\documentclass[%
 reprint,
 amsmath,amssymb,
 aps,
 prl,
]{revtex4-2}

\usepackage{graphicx,color}
\usepackage{dcolumn}
\usepackage{bm}
\usepackage{braket}
\usepackage[
colorlinks=true,
pdfstartview=FitV,
linkcolor=blue,
citecolor=magenta,
urlcolor=blue,
bookmarks=true,
bookmarksnumbered=true,
pdftitle={},
pdfauthor={}
]{hyperref}

\newcommand{\sectionprl}[1]{{\em #1}\/.---}

\begin{document}

\preprint{APS/123-QED}

\title{Kondo Effect in a Quantum Dot under Continuous Quantum Measurement}

\author{Masahiro Hasegawa$^{1}$, Masaya Nakagawa$^{2}$, and Keiji Saito$^{1}$}
\affiliation{$^{1}$Department of Physics, Keio university, Hiyoshi 3-14-1, Kohoku-ku, Yokohama, Japan}%
\affiliation{$^{2}$Department of Physics, University of Tokyo, 7-3-1 Hongo, Bunkyo-ku, Tokyo 113-0033, Japan}%

\date{\today}

\begin{abstract}
The backaction of quantum measurement on the Kondo effect in a quantum dot system is investigated by considering continuous projective measurement of singly occupied states of a quantum dot.
We elucidate the qualitative feature of the Kondo effect under quantum measurement and determine effective Kondo temperature affected by the measurement.
The Kondo resonance in the spectral function is suppressed when the measurement strength reaches the energy scale of the Kondo temperature without measurement.
Through the spin susceptibility, we identify the generalized Kondo temperature under continuous quantum measurement. 
The measurement backaction changes the singularity in the spin susceptibility into a highly non-monotonic temperature dependence around the generalized Kondo temperature.
The dependence of the generalized Kondo temperature on the measurement strength is quantitatively discussed.
\end{abstract}

\maketitle

\sectionprl{Introduction}
Quantum measurement is not only a unique property of itself that does not exist in classical mechanics, but it also leads to backaction on the physical object, producing several nontrivial phenomena \cite{wiseman2009quantum}.
Recent development in quantum technologies has facilitated the possibility of observation of measurement backaction from single quanta to many-body systems~\cite{Murch2008May,Patil2015Oct}.
For example, the continuous quantum Zeno effect has been observed in ultracold atoms subjected to strong particle loss~\cite{Syassen2008Jun,Barontini2013Jan,Zhu2014Feb,Tomita2017Dec}.
The measurement backaction naturally arises in dissipative open quantum systems since an environment may act on a system as an observer~\cite{Daley2009Jan,Guo2009Aug,Kantian2009Dec,Ruter2010Mar,Lee2014Oct,Zeuner2015Jul,Ashida2016Nov,Nakagawa2018Nov,Yamamoto2019Sep,Froml2019Feb,Wolff2020Feb,Froml2020Apr}.
This close relation between the measurement backaction and dissipation implies that new functionalities or novel transitions may be induced in quantum many-body systems using measurement backaction via dissipation engineering~\cite{Beige2000Aug,Kraus2008Oct,Diehl2009Nov,Verstraete2009Sep,Pedersen2014Nov,Li2018Nov,Skinner2019Jul,Lang2020Sep,Ippoliti2021Feb,Alberton2021Apr}. 

\begin{figure}[bth]
    \centering
    \includegraphics[width=0.9\linewidth]{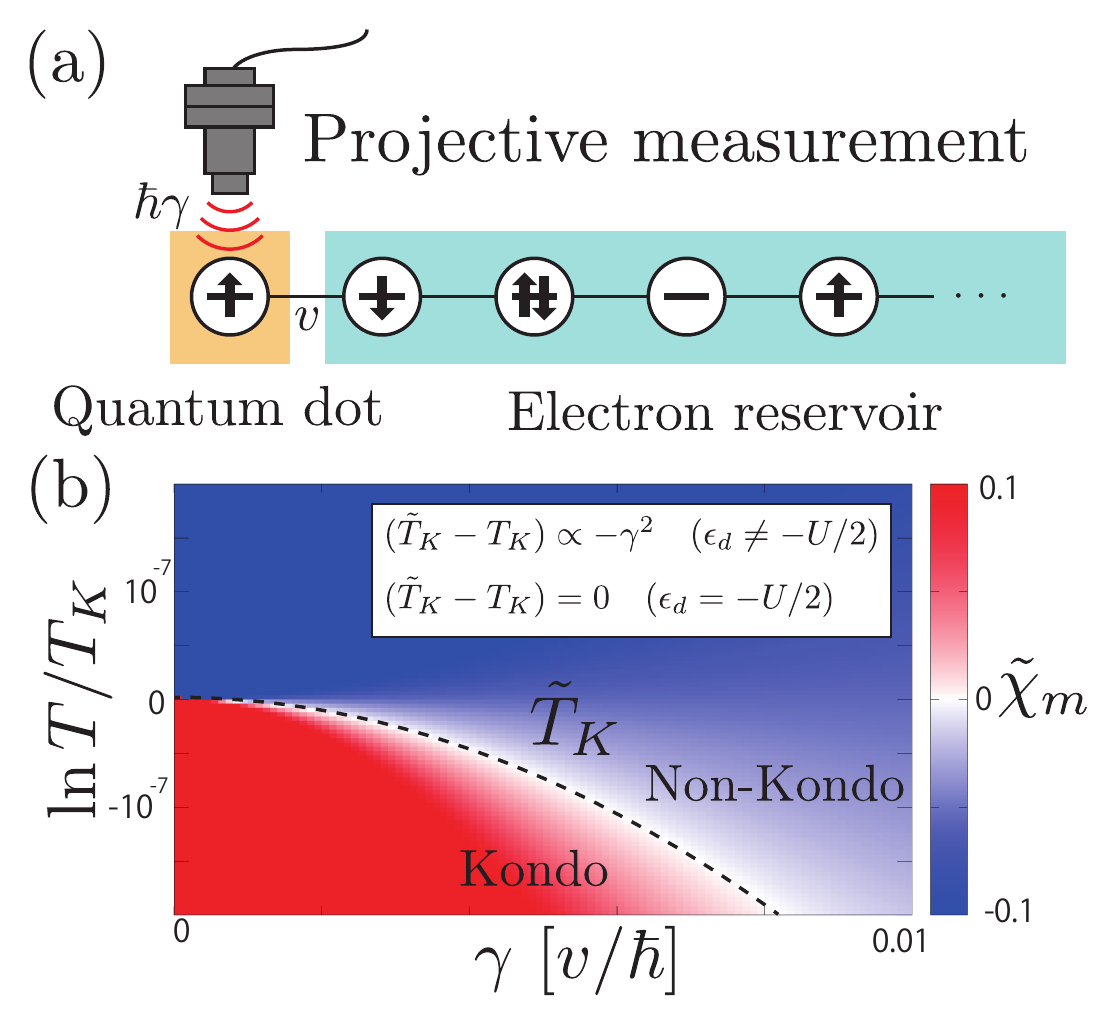}
    \caption{(a) Schematic of the considered system.
    A single-level quantum dot is connected to an electron reservoir with a tunnel coupling strength $v$, and a continuous projective measurement is applied to the dot with a coupling strength $\hbar \gamma$.
    (b) Schematic density plot of the singular part $\tilde{\chi}_m$ [Eq. (\ref{eqn:chi_tilde_Kondo})] of the spin susceptibility for $U/v = 10.0$ and $\epsilon_d/v = -3,0$.
    The horizontal axis denotes the strength of measurement, $\hbar \gamma / v$, and the vertical axis denotes the reservoir temperature $T$ in units of the Kondo temperature for $\gamma = 0$.
    The broken curve displays the generalized Kondo temperature defined in Eq. (\ref{eqn:generalized_Kondo_temperature_small}).
    The $\gamma$-dependence differs for the particle-hole symmetric and asymmetric cases.
    To highlight the gradient around the generalized Kondo temperature, we set the range of the color bar from $-0.1$  to $0.1$, though $\tilde{\chi}_m$ takes larger values around the diverging point.
    }
    \label{fig:system}
\end{figure}

The Kondo effect is a well-known nontrivial quantum many-body phenomenon in which a localized quantum spin forms a spin-singlet state with surrounding electrons~\cite{TextbookHewson1993}.
To observe quantum-measurement backaction in the Kondo effect, quantum dot systems are a promising platform~\cite{Goldhaber-Gordon1998Jan,Sakano2011Feb,Ferrier2016Mar,Hensgens2017Aug,Yoo2018Apr,Sivre2019Dec,Borzenets2020Mar,Hasegawa2021Jan}.
In quantum dot systems, many system parameters can be tuned, and the electronic state of the quantum dot can be continuously detected by real-time measurement techniques~\cite{Buks1998Feb,Wei2003May,Fujisawa2006Jun,Sukhorukov2007Mar}.
While dissipation engineering in quantum dot systems has been recently discussed~\cite{Ota2017Oct,Lourenco2018Aug,Damanet2019Oct,Liu2020Sep}, limited studies have been conducted on the effect of measurement backaction on the Kondo effect~\cite{Silva2003Apr,Avinun-Kalish2004Apr}.
Given the experimental developments in these decades, it is clearly fundamental to figure out the effects of quantum measurement on the Kondo effect.
In this paper, we step forward in this direction by considering a quantum dot under the projective measurement of singly occupied states of the dot. See Fig.~\ref{fig:system} (a).

Although our measurement setup is simple, its backaction on the Kondo effect is highly nontrivial since such projective measurement does not destroy a local magnetic moment inside the dot.
Therefore the Kondo singlet state survives under this measurement.
Note that the measurement detects the number of electrons, not the spin of the electrons.
Within this measurement framework, we extract general features of measurement backaction.
The main goals of this paper are as follows: capture the qualitative feature of the Kondo effect under quantum measurement, and determine effective Kondo temperature affected by the measurement.
To this end, we develop the Keldysh formalism that has recently been discussed for open quantum systems~\cite{Seiberer2016Aug}, together with the standard approaches for analyzing the pure Kondo effect~\cite{TextbookHewson1993}. In addition, we employ modified Schrieffer-Wolff transformation.
For the first aim, we calculate the spectral function of the quantum dot by the perturbation theory in the regime of weak Coulomb interaction. 
To achieve the second aim, we extend the Schrieffer-Wolff transformation to open quantum systems. Then we derive a generalized Kondo model under measurement, which involves a complex-valued spin-exchange coupling.
By calculating the spin susceptibility with the perturbation theory with respect to the complex-valued spin-exchange coupling, we found that the logarithmic singularity in the spin susceptibility is significantly altered due to the measurement backaction. 
The explicit dependence of the generalized Kondo temperature on the measurement strength is briefly summarized in Fig.~\ref{fig:system} (b).

\sectionprl{Setup}
We consider a quantum dot system under continuous projective measurement.
The quantum dynamics under continuous measurement is described by the following Lindblad equation after the ensemble average over measurement outcomes~\cite{TextbookCarmichael1993,Ashida2018}:
\begin{eqnarray}
    \frac{d}{dt} \rho &=& -\frac{i}{\hbar} [H,\rho] + \gamma \left( L \rho L^{\dagger} - \frac{1}{2} \{ L^{\dagger} L , \rho \} \right)\equiv\mathcal{L}\rho . \label{eqn:Lindblad_equation}
\end{eqnarray}
Here $\rho$ is a density matrix of the system and $\gamma$ is an amplitude (frequency) of the measurement.
$[\cdot,\cdot]$ and $\{ \cdot,\cdot \}$ are a commutator and an anti-commutator, respectively.
The Hamiltonian of the system is given by the Anderson impurity model:
\begin{eqnarray}
    H &=& \sum_s \left[ \epsilon_d d_{s}^{\dagger} d_s + \frac{U}{2} d_{s}^{\dagger} d_s d_{\bar{s}}^{\dagger} d_{\bar{s}} \right] + \sum_{k,s} \epsilon_{k}  c_{ks}^{\dagger} c_{ks} \nonumber  \\
    && + v \sum_{k,s} ( c_{ks}^{\dagger} d_s + d_{s}^{\dagger} c_{ks}) ,
\end{eqnarray}
where $d_s (d_s^{\dagger})$ is an annihilation (creation) operator of an electron in a quantum dot with spin $s\in\{\uparrow,\downarrow\}$ and $c_{ks} (c_{ks}^{\dagger})$ is that of an electron in an electron reservoir with wave number $k$ and spin $s$. The coefficients $U$, $\epsilon_d$, and $\epsilon_{k}$ are the Coulomb interaction strength, an energy level of the dot, and the energy dispersion of the reservoir, respectively, and $v$ is a tunnel coupling constant between the reservoir and the dot, where we assume 
the wide-band limit. Here $\bar{s}$ denotes the opposite component of spin $s$, i.e., $\bar{\uparrow} = \downarrow$. The jump operator $L$ is a projection $P$ onto the singly occupied subspace of the dot: 
\begin{eqnarray}
    L = P = \sum_s d^{\dagger}_s d_s d_{\bar{s}} d^{\dagger}_{\bar{s}} .
\end{eqnarray}
Hereafter we set $\hbar = k_B = 1$.

\sectionprl{Spectral function}
We first calculate the spectral function to obtain an indication of qualitative nature of the backaction on the Kondo effect.
The spectral function is defined as $\mathcal{A}_s(\omega) = \Gamma \,\mathrm{Im}[ G^A_s(\omega) ] /2 $, where $\Gamma = 2\pi \nu v^2$ is the linewidth with the density of the states of the reservoir $\nu$ and the function $G^A_s(\omega)$ is the Fourier transform of the advanced Green's function, $G^A_s(t) \equiv i\Theta(t) \Braket{ \{ d_s(0),d_s^{\dagger}(t) \} } $. Here 
$\langle ...\rangle$ implies a steady-state average.
The advanced Green's function can be expressed in the following form:
\begin{eqnarray}
    G^A_s(\omega) 
    &=& \frac{1}{\omega - \epsilon_d  - i \Gamma/2 - \Sigma_{s}^A(\omega)} ,
\end{eqnarray}
where $\Sigma_{s}^A(\omega)$ is the one-particle irreducible (1PI) self-energy. We calculate the self-energy within the second-order perturbation theory with respect to the Coulomb interaction strength~\cite{Hershfield1992Sep}, which is a minimal approximation for the Kondo resonance peak at the Fermi level in the particle-hole symmetric case ($\epsilon_d = - U/2$).

By using the Keldysh formalism for the Lindblad equation~\cite{Seiberer2016Aug}, we generalize the 1PI self-energy for nonzero $\gamma$. Up to the first order of $\gamma$, it is decomposed into three terms \cite{supple}:
\begin{eqnarray}
    \Sigma_{s}^A(\omega) = \Sigma_{s,U}^{(1),A}(\omega) + \Sigma_{s,\gamma}^{(1),A}(\omega) + \Sigma_{s,U}^{(2),A}(\omega) .
\end{eqnarray}
Here, $\Sigma_{s,U}^{(1),A}(\omega) = U n_{\bar{s}}$ and $\Sigma_{s,\gamma}^{(1),A}(\omega)= i \gamma / 2$ are the 1PI self-energies in the first order with respect to $U$ and $\gamma$, respectively, and $\Sigma_{s,U}^{(2),A}(\omega) $ is the 1PI self-energy in the second order with respect to $U$.
Here $n_s$ is the occupation number of the dot.
Within this form, the occupation number is calculated using the self-consistent equation:
\begin{eqnarray}
    n_s = \int \frac{d\omega}{2\pi} \frac{\Gamma f(\omega) + \gamma n_{s}}{(\omega-\epsilon_d - U n_{\bar{s}})^2 + (\Gamma+\gamma)^2/4} . \label{eqn:selfconsistent_1storder}
\end{eqnarray}
\begin{figure}[t]
    \centering
    \includegraphics[width=1.0\linewidth]{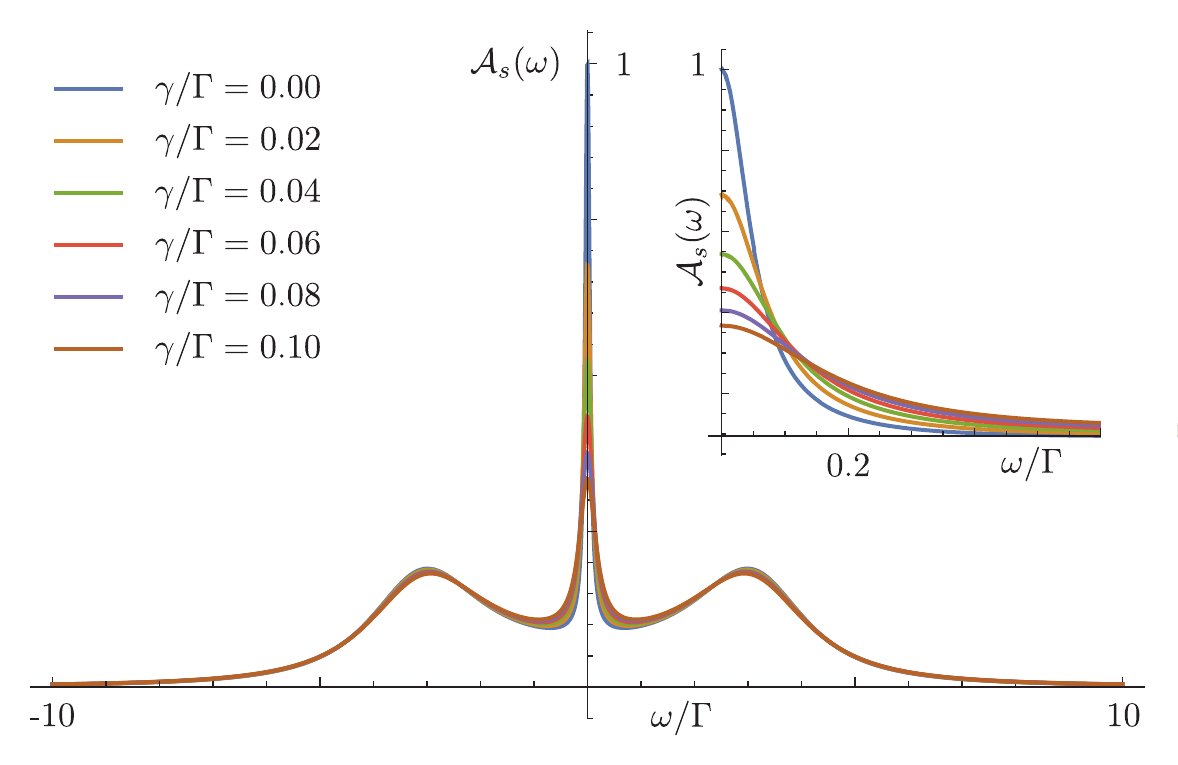}
    \caption{Frequency dependence of the spectral function $\mathcal{A}_s(\omega)$ for several values of $\gamma$.
    The parameters are set to $U = 6.0 \Gamma$ and $\epsilon_d = -3.0 \Gamma$. 
    The temperature of the electron reservoir is set to zero.
    The qualitative tendency of the suppression of the Kondo resonance peak is not changed by the finite-temperature effect \cite{supple}.
    The inset shows the spectral function near the Fermi level, $\omega = 0$.
    }
    \label{fig:spectral}
\end{figure}

Figure \ref{fig:spectral} displays the spectral function for several values of $\gamma$.
For $\gamma = 0$, the spectral function exhibits the Kondo resonance at the Fermi level, which indicates that the linear conductance reaches the unitary limit.
When the projective measurement is introduced, this Kondo resonance is suppressed and halved at $\gamma = \gamma_{\mathrm{half}} \sim 0.05 \Gamma$.
Note that $\gamma_{\mathrm{half}}$ should be of the same order of the Kondo temperature since the Kondo resonance peak is halved at the Kondo temperature, as inferred from the empirical formula~\cite{Goldhaber-Gordon1998Dec}.
In fact, $\gamma_{\mathrm{half}}$ and the Kondo temperature without measurement estimated from the Fermi liquid theory~\cite{Hewson2001Oct} are of the same order for the given parameters~\footnote{The Kondo temperature without measurement in the Fermi liquid theory is calculated as $T_K = \Gamma \sqrt{\pi u / 8 } e^{-\pi^2 u/8 + u/2}$, where $u= 2U/(\pi \Gamma)$. For $U=6.0\Gamma$, $T_K \sim 0.07 \Gamma$, which is of the order of $\gamma_{\mathrm{half}}$.}.

\sectionprl{Non-unitary Schrieffer-Wolff transformation}
The perturbation theory with respect to the Coulomb interaction strength is based on the weak-coupling approach in the spirit of the local Fermi liquid theory~\cite{TextbookHewson1993}.
In the opposite limit that describes the deep Kondo regime ($|v|\ll|\epsilon_d|,\epsilon_d+U$), a local moment is formed at the quantum dot, and the spin-exchange interaction between the local moment and the reservoir electrons plays the primary role in the Kondo effect.
To understand the effect of the measurement backaction in the deep Kondo regime, we generalize the Schrieffer-Wolff transformation (SWT)~\cite{Schrieffer1966Sep} to the Anderson model under continuous measurement.
The objective of using this transformation is to derive the effective Kondo Hamiltonian and then to generalize the Kondo temperature.

Here, we employ a non-unitary generalization of the SWT for the Liouvillian $\mathcal{L}$ defined in Eq.~\eqref{eqn:Lindblad_equation}~\cite{Kessler2012Jul,Rosso2020Dec}. 
We use this technique for weak measurement.
The non-unitary transformation requires meticulous treatment~\footnote{See also the supplemental material}.
As shown later, this method can be qualitatively justified by consistent results obtained for the weak measurement case by the unitary SWT.
In introducing the non-unitary SWT, it is convenient to use vector representation of the density matrix with a doubled Hilbert space as $\rho' :=\sum_{i_2,i_2} \rho_{i_1,i_2} \Ket{i_1} \otimes \Ket{i_2}$.
The Liouvillian $\mathcal{L}$ can similarly be identified with a linear operator $\mathcal{L}'$ acting on the doubled Hilbert space.
Then, the non-unitary SWT is defined by a transformation $\tilde{\mathcal{L}} = e^{S} \mathcal{L}' e^{-S}$ and $\tilde{\rho} = e^{S} \rho' $ with a non-Hermitian generator
\begin{eqnarray}
    S = S_1 \otimes Q + Q \otimes S_1^{\ast} + S_2 \otimes P + P \otimes S_2^{\ast} ,
\end{eqnarray}
\begin{eqnarray}
    S_i = \sum_{k,s} \sum_{\alpha = \{p,h \} } \left( \frac{v}{\epsilon_k - \epsilon_i^{\alpha}} n_{d\bar{s}}^{\alpha} c_{ks}^{\dagger} d_s  - \mathrm{h.c.} \right) , 
\end{eqnarray}
where $Q=1-P$, $n_{ds}^{p} = 1-n_{ds}^{h} = d_{s}^{\dagger} d_s$, $\epsilon_1^p = (\epsilon_2^p)^{\ast}  =\epsilon_d + U + i \gamma /2$, and $\epsilon_1^h = (\epsilon_2^h)^{\ast} = \epsilon_d - i \gamma /2$.
The non-unitary SWT decouples a subspace projected by $P \otimes P + Q \otimes Q$ from its complement up to the second order of the tunnel coupling.
We obtain an effective Liouvillian in the singly occupied subspace of the dot as \cite{supple}
\begin{eqnarray}
    \frac{d}{dt} \tilde{\rho}_{PP} = \tilde{\mathcal{L}}_{\mathrm{eff}} \tilde{\rho}_{PP}, \label{eqn:porjected_Lindblad}
\end{eqnarray}
where $\tilde{\rho}_{PP}=P \otimes P\tilde{\rho}$ and
\begin{eqnarray}
    i \tilde{\mathcal{L}}_{\mathrm{eff}} = P H_{\mathrm{eff}} P \otimes P - P \otimes P H_{\mathrm{eff}}^{\ast} P .
\end{eqnarray}
The effective Hamiltonian is given by a non-Hermitian Kondo model~\cite{Nakagawa2018Nov}
\begin{eqnarray}
    H_{\mathrm{eff}} &=& \sum_{k,s} \epsilon_{k} c_{ks}^{\dagger} c_{ks} + J \sum_{k_1,k_2} \bm{S}_{d} \cdot \bm{S}_{k_1k_2} , \label{eqn:effective_Hamiltonian}
\end{eqnarray}
where $\bm{S}_d$ and $\bm{S}_{k_1k_2}$ are the spin operators at the quantum dot and the electron reservoir, respectively.
The $i(=\!x,y,z)$th component of these operators are given as 
$S_{i,d} = (d_{\uparrow}^{\dagger},d_{\downarrow}^{\dagger}) (\sigma_i/2) (d_{\uparrow},d_{\downarrow})^T$ and $S_{i,k_1k_2} = (c_{k_1\uparrow}^{\dagger},c_{k_1\downarrow}^{\dagger}) (\sigma_i/2) (c_{k_2\uparrow},c_{k_2\downarrow})^T$ with the Pauli matrices $\sigma_i$.
In Eq.~\eqref{eqn:effective_Hamiltonian}, we drop nonessential  terms that are not related to the Kondo effect, such as a potential scattering term.
The coefficient $J$ is a complex-valued spin-exchange coupling strength at the Fermi level given by
\begin{eqnarray}
    J &=& -2 v^2 \Biggl( \frac{1}{\epsilon_d + i \gamma /2}  - \frac{1}{\epsilon_d + U- i \gamma /2}  \Biggr) . \label{eqn:complex_spin_exchange}
\end{eqnarray}
The dependence of the spin-exchange coupling on the wave numbers $k_1$ and $k_2$ is neglected since we focus on the scattering process around the Fermi level.
The complex-valued spin-exchange coupling may be interpreted as a consequence of the measurement backaction on intermediate states in the virtual processes in the second-order perturbation theory~\cite{Nakagawa2020Apr}.

\sectionprl{Spin susceptibility}
Having obtained the effective Kondo Hamiltonian (\ref{eqn:effective_Hamiltonian}), we now compute the spin susceptibility to identify the Kondo temperature.
Note that in the pure Kondo effect without measurement, the spin susceptibility exhibits a divergence at the Kondo temperature within the perturbation theory with respect to the spin-exchange coupling \cite{TextbookHewson1993}.
Here, we calculate the spin susceptibility of the quantum dot under continuous measurement by the perturbation theory with respect to the complex spin-exchange coupling $J$ and discuss how its singular behavior is altered by the measurement backaction.
The spin susceptibility is defined as $\chi_{m}(t) = (g \mu_B) (\partial \Braket{S_{z,d}(t)} / \partial B) |_{B=0}$, where $g$ and $\mu_B$ are the Land\'{e} g-factor and the Bohr magneton, respectively.
Here, the magnetic moment of the dot is coupled with the magnetic field $B$ through Zeeman coupling.

\begin{figure}
    \centering
    \includegraphics[width=1.0\linewidth]{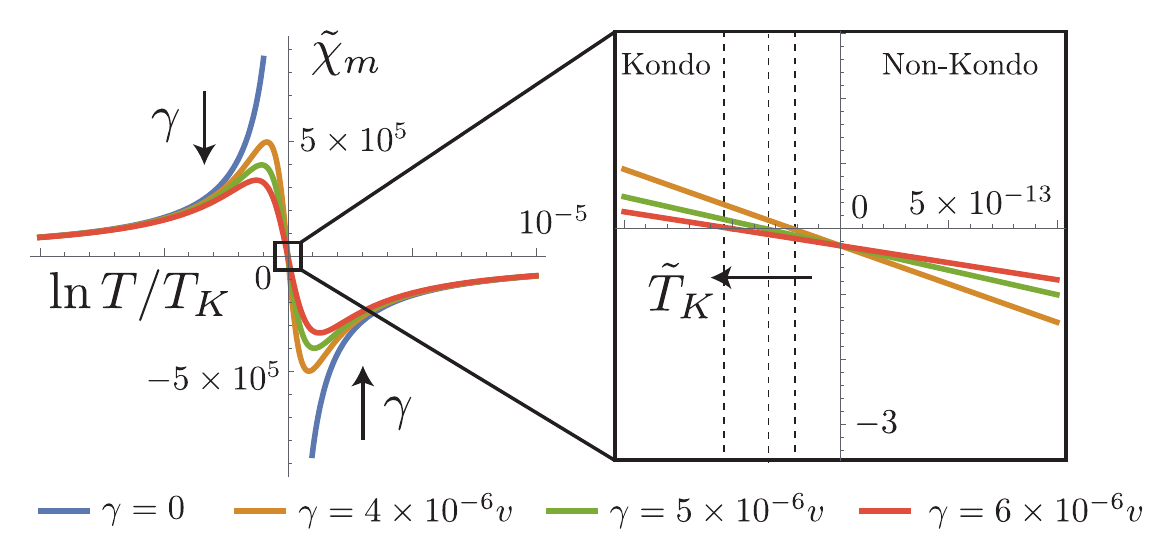}
    \caption{Temperature dependence of the singular contribution to the spin susceptibility around the Kondo temperature for $U/v=10.0$ and $\epsilon_d/v = -3.0$.
    The divergence at the Kondo temperature is replaced by a zero crossing at the generalized Kondo temperature for nonzero $\gamma$.
    The generalized Kondo temperature decreases as $\gamma$ increases in the particle-hole asymmetric case.
    }
    \label{fig:tildechim}
\end{figure}

Using the Keldysh formalism for the Lindblad equation~\cite{Seiberer2016Aug,Froml2019Feb,Froml2020Apr}, we compute the spin susceptibility of a steady state. By summing up the contribution whose Green's functions of electrons in the reservoir are connected in a single loop and all lined up on either the forward time contour or the backward time contour \cite{supple}, 
we extract a leading singular contribution to the spin susceptibility as
\begin{eqnarray}
    \tilde{\chi}_{m} \simeq - \frac{\chi_{m,0}}{\sqrt{3} a} \mathrm{Re} \left[ \frac{a J \nu}{1 + a J \nu \ln T / D} \right] . \label{eqn:chi_tilde_Kondo}
\end{eqnarray}
Here, $\chi_{m,0}$ is the spin susceptibility for $v=0$, $D$ is the bandwidth of the reservoir, and $a = (1+\sqrt{3})/4$.
In the $\gamma \to 0$ limit, $\tilde{\chi}_{m}$ exhibits a divergence at the Kondo temperature $T = T_K = D e^{-(a J \nu)^{-1}}$.
For nonzero $\gamma$, the divergence at the Kondo temperature reduces to a zero crossing at a generalized Kondo temperature 
\begin{eqnarray}
    \tilde{T}_K &=&  D \exp \left[- \frac{1}{a\nu} \frac{\mathrm{Re}J}{(\mathrm{Re}J)^2 + (\mathrm{Im}J)^2} \right] \nonumber \\
    &=& T_K \left[ 1 - \frac{\pi}{4} \frac{(U+2\epsilon_d)^2}{a \Gamma U^3} \gamma^2 \right] + O(\gamma^4) \leq T_K , \label{eqn:generalized_Kondo_temperature_small} 
\end{eqnarray}
and $\tilde{\chi}_{m}$ exhibits a highly non-monotonic temperature dependence around $\tilde{T}_K$ (see Fig.~\ref{fig:tildechim}).
We present a schematic density plot of $\tilde{\chi}_{m}$ in Fig. \ref{fig:system} (b), where the generalized Kondo temperature is shown by the broken curve.
We note that $\tilde{T}_K=T_K$ holds for $\epsilon_d = -U/2$; thus, the Kondo temperature is unaffected by the measurement backaction in the particle-hole symmetric case.

\sectionprl{Interpretation of the generalized Kondo temperature}
The qualitative behavior of the Kondo effect under continuous quantum measurement can also be understood from the standard unitary SWT on the Anderson impurity model.
When the unitary SWT is applied to the Lindblad equation~[Eq. \eqref{eqn:Lindblad_equation}], the states projected by the jump operator are no longer singly occupied states of the dot in the Kondo model.
For example, a local spin-singlet state in the Anderson impurity model is transformed into a superposition of four states $\Ket{\uparrow \downarrow}_{\mathrm{K}}, \Ket{\downarrow \uparrow}_{\mathrm{K}} ,\Ket{02}_{\mathrm{K}}$, and $\Ket{20}_{\mathrm{K}}$ \cite{supple}.
Here, $\ket{\sigma}$, $\ket{0}$, and $\ket{2}$ denote a singly occupied state with spin $\sigma$, an empty state, and a doubly occupied state, respectively, and $\Ket{ij}_{\mathrm{K}}$ denotes a quantum state in the frame after the SWT where the dot site is in $\Ket{i}$ state and the reservoir site next to the dot is in $\Ket{j}$ state.

To see the effect of the projective measurement of this hybridized state, we consider the time evolution of a singly occupied state $\rho = (\Ket{\uparrow \downarrow} \Bra{\uparrow \downarrow}_{\mathrm{K}} + \Ket{\downarrow \uparrow } \Bra{\downarrow \uparrow }_{\mathrm{K}} )/2$.
The $\gamma$-dependent diagonal terms of the density matrix are calculated from the Lindblad equation as \cite{supple}
\begin{eqnarray}
    \Bra{\uparrow \downarrow} \dot{\rho} \Ket{\uparrow \downarrow}_{\mathrm{K}} &=& \Bra{\downarrow \uparrow} \dot{\rho} \Ket{\downarrow \uparrow }_K \nonumber \\
    &\sim & - \frac{\gamma v^2}{2} \left[ \frac{1}{\epsilon_d^2} + \frac{1}{(\epsilon_d+U)^2} \right] < 0 , \nonumber \\
    \Bra{20} \dot{\rho} \Ket{20}_{\mathrm{K}} &\sim&  \gamma v^2 \frac{1}{(\epsilon_d+U)^2} > 0, \nonumber \\
    \Bra{02} \dot{\rho} \Ket{02}_{\mathrm{K}} &\sim&  \gamma v^2 \frac{1}{\epsilon_d^2} > 0 . \label{eqn:time_evolution_diagonal}
\end{eqnarray}
Thus, the projective measurement of the hybridized state decreases the probability distribution of the singly occupied states and facilitates charge fluctuations, thereby effectively weakening the singularity in the Kondo effect.

The $\gamma$-dependence of the generalized Kondo temperature in Eq.~\eqref{eqn:generalized_Kondo_temperature_small} is consistent with the above picture.
The decrease rate of the probability in the singlet state in Eq. (\ref{eqn:time_evolution_diagonal}) takes its minimum at $\epsilon_d = - U/2$ under fixed $J$.
Since the outflow of the probability distribution of the singlet states is suppressed for $\epsilon_d = - U/2$, the generalized Kondo temperature \eqref{eqn:generalized_Kondo_temperature_small} takes its maximum value in the particle-hole symmetric case.

\sectionprl{Summary and discussion}
To elucidate the effect of measurement backaction on quantum many-body physics, we have analyzed the Kondo effect in a quantum dot under continuous projective measurement of singly occupied states of the dot. 
The main results are summarized in Fig.\ref{fig:system} (b). 
The result of the spectral function has revealed that the measurement backaction suppresses the Kondo resonance peak when the strength of the measurement is of the order of the Kondo temperature.
Furthermore, we have shown that the divergent behavior of the spin susceptibility near the Kondo temperature is replaced by a zero crossing with highly non-monotonic temperature dependence that is characterized by the generalized Kondo temperature.

The generalized Kondo temperature in Eq.~\eqref{eqn:generalized_Kondo_temperature_small} is consistent with the energy scale of a circular renormalization-group flow of the non-Hermitian Kondo model \eqref{eqn:effective_Hamiltonian} in Ref.~\onlinecite{Nakagawa2018Nov}~\footnote{In Ref.~\onlinecite{Nakagawa2018Nov}, a two-loop renormalization-group analysis has shown a quantum phase transition between the Kondo phase and the non-Kondo phase due to non-Hermiticity. The present result is consistent with a one-loop renormalization-group flow in which the transition occurs for an infinitesimal imaginary part of $J$.}.
While the signature of the new energy scale has remained elusive in Ref.~\onlinecite{Nakagawa2018Nov}, our result derived from the spin susceptibility shows that the generalized Kondo temperature characterizes the non-monotonic temperature dependence of physical quantities in a quantum dot under continuous measurement.

Our result in Eq. (\ref{eqn:chi_tilde_Kondo}) reveals that the singular behavior of physical quantities in the Kondo problem is significantly altered due to measurement backaction. Nevertheless, we note that Eq. (\ref{eqn:chi_tilde_Kondo}) is not an exact susceptibility and only extracts the singularity from specific contributions.
To calculate the spin susceptibility in the low-temperature Kondo regime, developing reliable numerical methods for the Kondo effect in open quantum systems is desired.

In quantum dot systems, real-time detection techniques for electronic states have been developed experimentally~\cite{Buks1998Feb,Wei2003May,Avinun-Kalish2004Apr,Fujisawa2006Jun,Sukhorukov2007Mar}.
We hope that further development of detection techniques in the near future would finally enable us to observe quantum measurement backaction on Kondo phenomena.

\section*{Acknowledgments}
The authors thank K. Kobayashi for the fruitful comments about the experimental realization.
M.H. was supported by Grants-in-Aid for Scientific Research JP19H05603.
M.N. was supported by Grants-in-Aid for Scientific Research JP20K14383.
K.S. was supported by Grants-in-Aid for Scientific Research (JP19H05603, JP19H05791).

\bibliography{references}

\end{document}


\preprint{APS/123-QED}

\title{Supplemental Material for \\ ``Kondo Effect in a Quantum Dot under Continuous Quantum Measurement"}

\author{Masahiro Hasegawa$^1$, Masaya Nakagawa$^2$, and Keiji Saito$^1$}
\affiliation{$^1$Department of Physics, Keio university, Hiyoshi 3-14-1, Kohoku-ku, Yokohama, Japan}%
\affiliation{$^2$Department of Physics, University of Tokyo, 7-3-1 Hongo, Bunkyo-ku, Tokyo 113-0033, Japan}%

\date{\today}


\maketitle

\section{Details of the perturbation theory with respect to the Coulomb interaction strength and the measurement strength}

In this section, we consider the perturbation theory with respect to the Coulomb interaction strength $U$ and the measurement strength $\gamma$.
Within this perturbation scheme, arbitrary strength of the hybridization $v_{k}$ is considered, while $U$ and $\gamma$ should be small.

Here, we consider the following Anderson impurity model:
\begin{eqnarray}
    H = H_0 + H_1,
\end{eqnarray}
where
\begin{eqnarray}
    H_0 &=& \sum_s \epsilon_d d_s^{\dagger} d_s + \sum_{k,s} \epsilon_k c_{ks}^{\dagger} c_{ks} , \\
    H_1 &=& \sum_s \frac{U}{2} d_{s}^{\dagger}d_s d_{\bar{s}}^{\dagger}d_{\bar{s}}  +  \sum_{k,s} v_k ( c_{ks}^{\dagger} d_s + d_s^{\dagger} c_{ks} ) .
\end{eqnarray}
Following Ref.~\cite{Seiberer2016Aug}, 
we define the Keldysh action as
\begin{eqnarray}
    \mathcal{S} = \mathcal{S}_0 + \mathcal{S}_1 + \mathcal{S}_2 ,
\end{eqnarray}
where 
\begin{eqnarray}
    \mathcal{S}_0 &=& \int_{K} d t \ \Biggl\{ \sum_s d_s^{\dagger}(t) i \partial_t d_s(t)  +  \sum_{k,s} c_{ks}^{\dagger}(t) i \partial_t c_{ks}(t)  -  H_0(t) \Biggr\} , \label{eqn:action_non_perturbed_U} \\
    \mathcal{S}_1 &=& - \int_R dt \  H_1(t) , \\
    \mathcal{S}_2 &=& - i \gamma \int_{t_i}^{t_f} dt \ [ L(t_+) L^{\dagger}(t_-)  - \frac{1}{2} L^{\dagger}(t_+) L(t_+) - \frac{1}{2} L^{\dagger}(t_-) L(t_-) ] .
\end{eqnarray}
The integrals $\int_K dt$ and $\int_R dt$ are performed on the Keldysh-Schwinger path and defined as follows:
\begin{eqnarray}
    \int_K dt \ f(t) &=& \int_{t_i}^{t_f} dt_+ \ f(t_+) -  \int_{t_i}^{t_f} dt_- \ f(t_-)  + \int_{t_i}^{t_i-i\beta} dt_M \ f(t_M) ,
\end{eqnarray}
and
\begin{eqnarray}
    \int_R dt \ f(t) = \int_{t_i}^{t_f} dt_+ \  f(t_+) -  \int_{t_i}^{t_f} dt_- \  f(t_-) ,
\end{eqnarray}
where $t_i$ and $t_f$ denote the initial time and the final time, respectively.
We focus on the steady state by taking $t_f = - t_i = \infty$.
Since the steady state does not depend on the initial condition, for convenience, we assume that the initial state is a thermal equilibrium state with inverse temperature $\beta$ in the limit of $v_k = U = \gamma =0$, and the interactions are introduced at the initial time $t_i$.

\subsection{Green's function}

The Green's function (GF) of electrons in the quantum dot is defined as
\begin{eqnarray}
    G_s^{\mu_1\mu_2}(t_1,t_2) = (-i) \Braket{\mathcal{T}_K d_s(t_{1\mu_1}) d_{s}^{\dagger}(t_{2\mu_2}) } ,
\end{eqnarray}
where $\mathcal{T}_K$ denotes the path ordering, and the Keldysh index $\mu_{i} = \pm$ indicates that the time variable $t_{i\mu_i}$ is on the forward (backward) contour for $\mu_{i} = +$ ($\mu_{i} = -$).

The perturbation theory consists of two steps.
First, we calculate the GF by expanding it with respect to the tunnel coupling $v_k$ in the non-interacting limit ($U = \gamma = 0 $).
The GF $G_{s,0}^{\mu_1\mu_2}(t_1,t_2)$ in the non-interacting limit satisfies the Dyson equation as follows:
\begin{eqnarray}
    G_{s,0}^{\mu_1\mu_2}(t_1,t_2) &=&  g_{s}^{\mu_1\mu_2}(t_1,t_2)  \nonumber \\
    &&+ \sum_{\mu_3,\mu_4} (-1)^{\mu_3 \mu_4} \int_{t_i}^{t_f} dt_3 dt_4 \ g_{s}^{\mu_1\mu_3}(t_1,t_3)  \Sigma_{s,v}^{\mu_3\mu_4}(t_3,t_4) G_{s,0}^{\mu_4\mu_2}(t_4,t_2) , \label{eqn:Dyson_nonint}
\end{eqnarray}
where $g_{s}^{\mu_1\mu_2}(t_1,t_2) $ is the GF for $v_k = U = \gamma =0$ and
$(-1)^{\mu_3 \mu_4}$ is a sign of the Keldysh time integral defined as $(-1)^{++} = (-1)^{--} = -(-1)^{+-} = -(-1)^{-+} = 1$.
Here, $\Sigma_{s,v}^{\mu_3\mu_4}(t_3,t_4)$ is a one-particle-irreducible (1PI) self-energy due to tunnel coupling.
Because the Hamiltonian of the tunnel coupling is quadratic, Eq. (\ref{eqn:Dyson_nonint}) is soluble for an arbitrary $v_k$.
In the wide-band limit, the momentum dependence of the tunnel coupling is neglected (i.e., $v_k = v$), and the GFs are calculated as
\begin{eqnarray}
    G_{s,0}^R(t_1,t_2) = [G_{s,0}^A(t_2,t_1)]^{\ast} = \int \frac{d\omega}{2\pi} e^{i \omega (t_1-t_2)} \frac{1}{\omega-\epsilon_d + i \Gamma/2}  ,
\end{eqnarray}
\begin{eqnarray}
    G_{s,0}^{+-}(t_1,t_2) = i \int \frac{d\omega}{2\pi} e^{i \omega (t_1-t_2)} \frac{\Gamma}{(\omega-\epsilon_d)^2 + \Gamma^2/4} f(\omega)  , 
\end{eqnarray}
where $\Gamma = 2\pi \nu v^2$ is the linewidth, $\nu$ is the density of states at the Fermi level, and $f(\omega) = [1 + e^{\beta \omega}]^{-1}$ is the Fermi distribution function of the electron reservoir.
The superscripts ${\cdot}^R$ and ${\cdot}^A$ denote the retarded and advanced components, respectively, defined as ${\cdot}^R = {\cdot}^{++} - {\cdot}^{+-}$ and ${\cdot}^A = {\cdot}^{++} - {\cdot}^{-+}$.

The next step involves performing the perturbation with respect to $U$ and $\gamma$.
Since the tunnel coupling Hamiltonian is quadratic, we can apply the Wick's theorem and the Dyson equation for the GF $G_s^{\mu_1\mu_2}(t_1,t_2)$ in the interacting case reads
\begin{eqnarray}
    G_s^{\mu_1\mu_2}(t_1,t_2) &=&  G_{s,0}^{\mu_1\mu_2}(t_1,t_2) \nonumber \\
    & & + \sum_{\mu_3,\mu_4} (-1)^{\mu_3 \mu_4} \int_{t_i}^{t_f} dt_3 dt_4 \ G_{s,0}^{\mu_1\mu_3}(t_1,t_3)  \Sigma_s^{\mu_3\mu_4}(t_3,t_4) G_s^{\mu_4\mu_2}(t_4,t_2) ,
\end{eqnarray}
where $\Sigma_s^{\mu_3\mu_4}(t_3,t_4)$ is the 1PI self-energy due to the Coulomb interaction and measurement backaction.
We calculate the 1PI self-energy up to the second order of $U$ and the first order of $\gamma$.
The self-energy is categorized into the following three parts:
a $\gamma$-independent part,
\begin{eqnarray}
    \Sigma_{s,U}^{\mu_3\mu_4}(t_3,t_4) & = & \lim_{\gamma \to 0}\Sigma_s^{\mu_3\mu_4}(t_3,t_4) ,
\end{eqnarray}
a $U$-independent part,
\begin{eqnarray}
    \Sigma_{s,\gamma}^{\mu_3\mu_4}(t_3,t_4) & = & \lim_{U \to 0}\Sigma_s^{\mu_3\mu_4}(t_3,t_4) ,
\end{eqnarray}
and a composite part,
\begin{eqnarray}
    \Sigma_{s,U\gamma}^{\mu_3\mu_4}(t_3,t_4) & = & \Sigma_s^{\mu_3\mu_4}(t_3,t_4) - \Sigma_{s,U}^{\mu_3\mu_4}(t_3,t_4) - \Sigma_{s,\gamma}^{\mu_3\mu_4}(t_3,t_4). 
\end{eqnarray}

\subsection{Self-consistent first-order approximation}
\begin{figure}[t]
    \centering
    \includegraphics[width=1.0\linewidth]{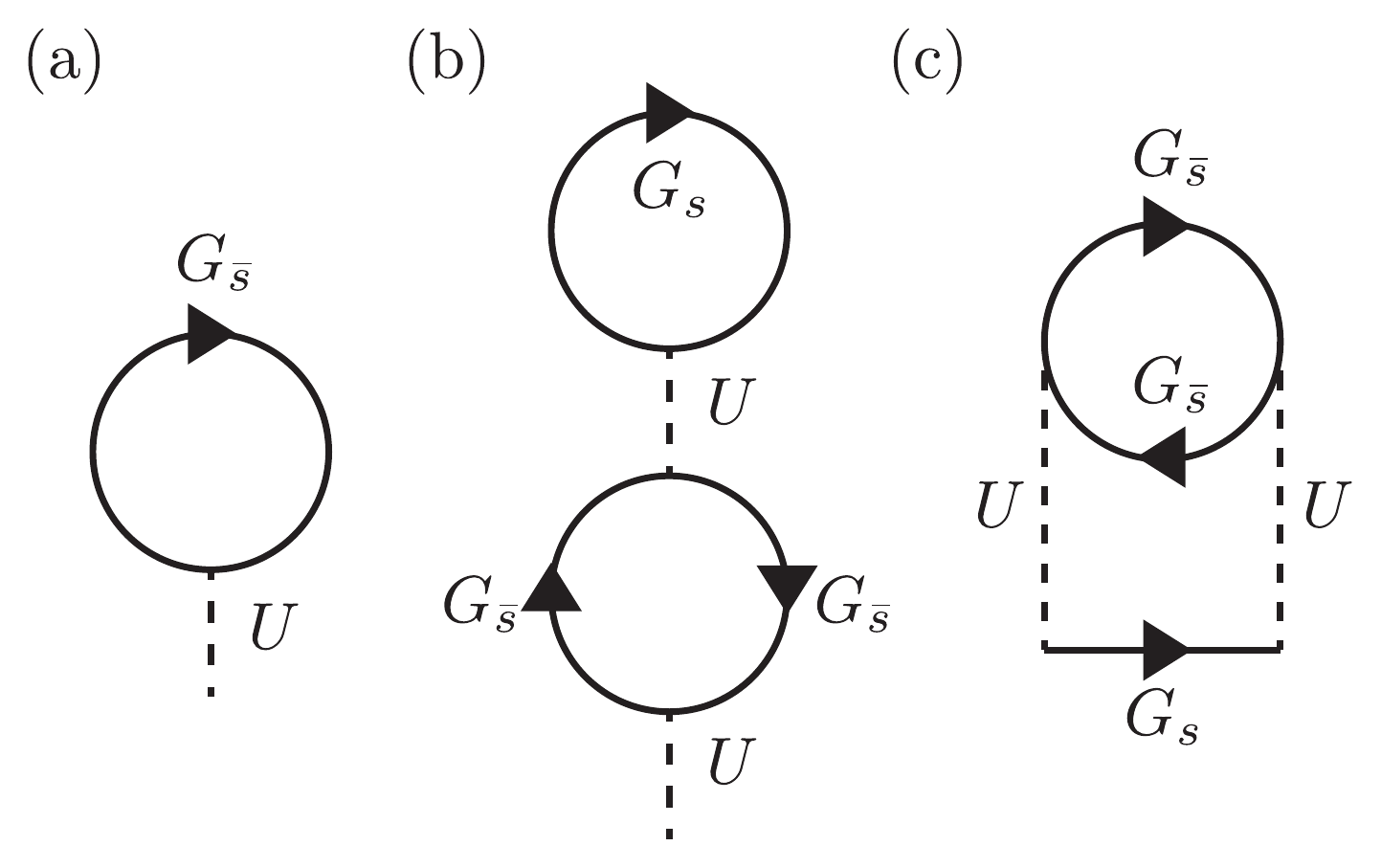}
    \caption{Feynman diagrams for the calculation of the self-energy due to the Coulomb interaction.}
    \label{fig:diagramsU}
\end{figure}

For the lowest-order calculation, we consider the self-consistent first-order approximation.
Without the measurement ($\gamma = 0$), this procedure corresponds to the Hartree approximation.
In this approximation, the self-energies are expressed as follows:
\begin{eqnarray}
    \Sigma_{s,U}^{\mu_1 \mu_2}(t_1,t_2) &=& \Sigma_{s,U}^{(1),\mu_1 \mu_2}(t_1,t_2) , \nonumber \\
    \Sigma_{s,\gamma}^{\mu_1 \mu_2}(t_1,t_2) &=& \Sigma_{s,\gamma}^{(1),\mu_1 \mu_2}(t_1,t_2) , \nonumber \\
    \Sigma_{s,U\gamma}^{\mu_1 \mu_2}(t_1,t_2) &=& 0 . 
\end{eqnarray}
Within the first-order approximation, the composite part of the self-energy vanishes.
Here, $\Sigma_{s,U}^{(1),\mu_1 \mu_2}(t_1,t_2)$ is the self-energy due to the Coulomb interaction in the first order (see Fig.~\ref{fig:diagramsU} (a) for the corresponding Feynman diagram), and calculated as
\begin{eqnarray}
    \Sigma_{s,U}^{(1),++}(t_1,t_2) = -\Sigma_{s,U}^{(1),--}(t_1,t_2) =  U n_{\bar{s}} \delta(t_1-t_2)  , \label{eqn:selfenergy_U_1st_ppmm}
\end{eqnarray}
\begin{eqnarray}
    \Sigma_{s,U}^{(1),+-}(t_1,t_2) = \Sigma_{s,U}^{(1),-+}(t_1,t_2) = 0 ,
\end{eqnarray}
\begin{eqnarray}
    \Sigma_{s,U}^{(1),R}(t_1,t_2) = \Sigma_{s,U}^{(1),A}(t_1,t_2) =  U n_{\bar{s}} \delta(t_1-t_2) ,
\end{eqnarray}
where $n_{s} = -i G_{s}^{+-}(t,t)$ is the expectation value of the occupation number of the dot.
$\Sigma_{s,\gamma}^{(1),\mu_1 \mu_2}(t_1,t_2)$ is the self-energy due to the measurement backaction in the first order, that is calculated as
\begin{eqnarray}
    \Sigma_{s,\gamma}^{(1),++}(t_1,t_2) = \Sigma_{s,\gamma}^{(1),--}(t_1,t_2) = i \gamma \frac{ 2 n_{s} - 1}{2} \delta(t_1-t_2) , \label{eqn:selfenergy_gamma_1st_ppmm}
\end{eqnarray}
\begin{eqnarray}
    \Sigma_{s,\gamma}^{(1),+-}(t_1,t_2) &=&  i \gamma n_{s} \delta(t_1-t_2) ,
\end{eqnarray}
\begin{eqnarray}
    \Sigma_{s,\gamma}^{(1),-+}(t_1,t_2) &=&  i \gamma (n_{s} -1) \delta(t_1-t_2) ,
\end{eqnarray}
\begin{eqnarray}
    \Sigma_{s,\gamma}^{(1),R}(t_1,t_2) = [ \Sigma_{s,\gamma}^{(1),A}(t_1,t_2) ]^{\ast} = - i \frac{\gamma}{2} \delta(t_1-t_2) . \label{eqn:selfenergy_gamma_1st_RA}
\end{eqnarray}
See the next section for details of the derivation of Eqs.~\eqref{eqn:selfenergy_gamma_1st_ppmm}-\eqref{eqn:selfenergy_gamma_1st_RA}.
Using Eqs.~\eqref{eqn:selfenergy_U_1st_ppmm}-\eqref{eqn:selfenergy_gamma_1st_RA}, we obtain the GFs within the first-order approximation as
\begin{eqnarray}
    G_{s}^{(1),R}(t_1,t_2) = [G_{s}^{(1),A}(t_1,t_2)]^{\ast} = \int \frac{d\omega}{2\pi} e^{i\omega(t_1-t_2)} \frac{1}{\omega - \epsilon_d - U n_{\bar{s}} + i (\Gamma+\gamma)/2}  ,  \label{eqn:GF_1stapp_RA}
\end{eqnarray}
\begin{eqnarray}
    G_s^{(1),+-}(t_1,t_2) = i \int \frac{d\omega}{2\pi} e^{i\omega(t_1-t_2)} \frac{\Gamma f(\omega) + \gamma n_s}{(\omega - \epsilon_d - U n_{\bar{s}})^2 + (\Gamma+\gamma)^2 /4} .  \label{eqn:GF_1stapp_les}
\end{eqnarray}

The GFs in Eqs. (\ref{eqn:GF_1stapp_RA}) and (\ref{eqn:GF_1stapp_les}) depend on $n_{s}$, which is determined from a self-consistent equation for the occupation number
\begin{eqnarray}
    n_s = \int \frac{d\omega}{2\pi} \frac{\Gamma f(\omega) + \gamma n_{s}}{(\omega-\epsilon_d - U n_{\bar{s}})^2 + (\Gamma+\gamma)^2/4} . \label{eqn:selfconsistent_1storder}
\end{eqnarray}
Equation (\ref{eqn:GF_1stapp_RA}) displays that the measurement backaction effectively decreases the lifetime of the dot state.

\subsection{Second-order approximation}

\begin{figure}[t]
    \centering
    \includegraphics[width=0.8\linewidth]{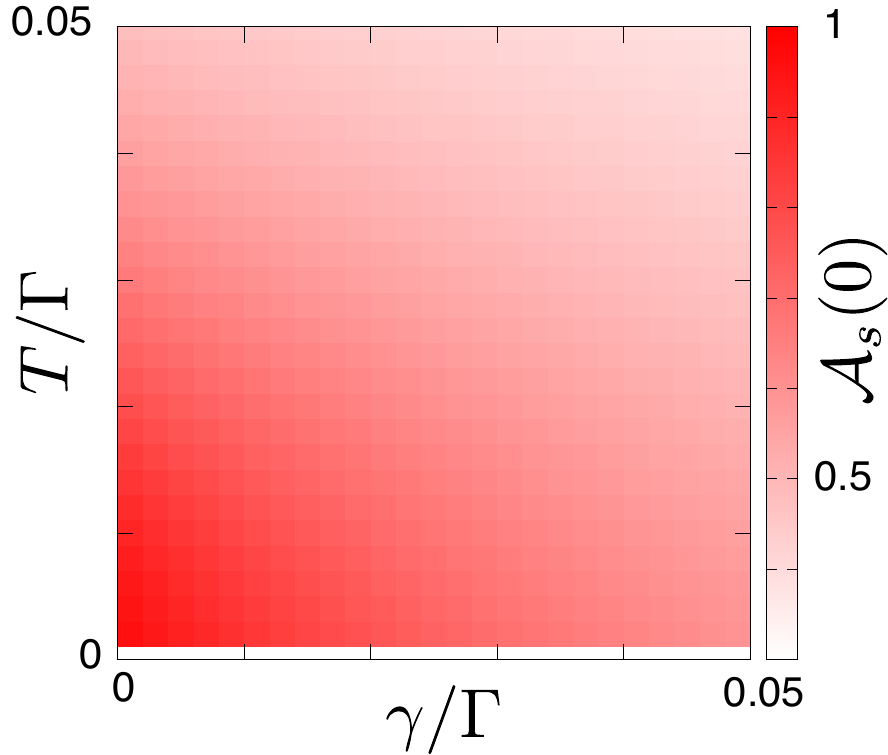}
    \caption{Density plot of the spectral function $\mathcal{A}_s(0)$ at the Fermi level.
    The horizontal and the vertical axes denote $\gamma / \Gamma$ and $T / \Gamma$, respectively.
    The Coulomb interaction strength and energy level of the quantum dot are set to $U = 6.0 \Gamma$ and $\epsilon_d = -3.0 \Gamma$, respectively.
    }
    \label{fig:spect_density}
\end{figure}

The minimal approximation that can describe the Kondo behavior is the second-order approximation~\cite{Hershfield1992Sep}, within which the Kondo resonance peak is developed at the Fermi level in the particle-hole symmetric case ($\epsilon_d = - U/2$).
In the second-order approximation, the self-energy is given by 
\begin{eqnarray}
    \Sigma_{s,U}^{\mu_1 \mu_2}(t_1,t_2) &=& \Sigma_{s,U}^{(2),\mu_1 \mu_2}(t_1,t_2) ,  
\end{eqnarray}
where
\begin{eqnarray}
    \Sigma_{s,U}^{(2),\mu_1 \mu_2}(t_1,t_2)= U^2 G_s^{(1),\mu_1 \mu_2}(t_1,t_2) G_{\bar{s}}^{(1),\mu_1\mu_2}(t_1,t_2) G_{\bar{s}}^{(1),\mu_2\mu_1}(t_2,t_1) . \label{eqn:self-energy_2ndorder}
\end{eqnarray}
This contribution corresponds to the Feynman diagram shown in Fig. \ref{fig:diagramsU} (c).
Note that we have replaced the GFs in Eq.~\eqref{eqn:self-energy_2ndorder} with those in the first-order approximation. Thus, the contribution from the Feynman diagrams in the first-order approximation is already included in the GFs, and it is not necessary to consider tadpole-type diagrams such as those shown in Fig. \ref{fig:diagramsU} (b).
Substituting the self-energy [Eq.~(\ref{eqn:self-energy_2ndorder})] into the GFs [Eqs. (\ref{eqn:GF_1stapp_RA}) and (\ref{eqn:GF_1stapp_les})], the spectral function presented in the main text can be calculated.

In the main text, we only discussed the spectral function at zero temperature.
Here, we discuss the finite-temperature effect on the spectral function.
Figure \ref{fig:spect_density} displays the spectral function at the Fermi level for several values of temperature and measurement strength.
For low temperature and weak measurement strength, the spectral function exhibits a strong Kondo resonance.
As the temperature and measurement strength increase, the spectral function at the Fermi level decreases.
This result can be understood from Eq.~\eqref{eqn:GF_1stapp_les}, where the numerator of the integrand reveals that the contribution from the measurement backaction can be regarded as a coupling to an effective infinite-temperature bath.

\section{Self-energy due to the measurement backaction}

Here, we calculate the self-energy due to the measurement backaction shown in Eqs. (\ref{eqn:selfenergy_gamma_1st_ppmm} - \ref{eqn:selfenergy_gamma_1st_RA}).
Before discussing the details of the calculation, we remark on the ordering of the creation and annihilation operators included in the jump operator $L$ because the jump operator is not written in the normal order.
To maintain the correct ordering, we introduce infinitesimal time differences to the jump operator as
\begin{eqnarray}
    L^{\dagger}(t) L(t) &=& \sum_{s_1,s_2} d_{s_1}^{\dagger}(t_1) d_{s_1}(t_2)  d_{\bar{s}_1}(t_3)  d_{\bar{s}_1}^{\dagger}(t_4)  d_{s_2}^{\dagger}(t_5) d_{s_2}(t_6)  d_{\bar{s}_2}(t_7)  d_{\bar{s}_2}^{\dagger}(t_8) ,
\end{eqnarray}
where the time variables $t_i$ are shifted from $t$ by infinitesimals so that $t_1 > t_2 > \cdots > t_8$ is satisfied.
With this jump operator, the first-order contribution to the GF $G_s^{\mu_1 \mu_2}(\tau_1,\tau_2)$ with respect to $\gamma$ is given by
\begin{eqnarray}
    && - (-i) \frac{\gamma}{2} \int dt \sum_{s_1,s_2} \Braket{\mathcal{T}_K d_s(\tau_1) d_s^{\dagger}(\tau_2)  d_{s_1}^{\dagger}(t_{1+}) d_{s_1}(t_{2+})  d_{\bar{s}_1}(t_{3+})  d_{\bar{s}_1}^{\dagger}(t_{4+})  d_{s_2}^{\dagger}(t_{5+}) d_{s_2}(t_{6+})  d_{\bar{s}_2}(t_{7+})  d_{\bar{s}_2}^{\dagger}(t_{8+}) } \nonumber \\
    && - (-i) \frac{\gamma}{2} \int dt \sum_{s_1,s_2} \Braket{\mathcal{T}_K d_s(\tau_1) d_s^{\dagger}(\tau_2)  d_{s_1}^{\dagger}(t_{1-}) d_{s_1}(t_{2-})  d_{\bar{s}_1}(t_{3-})  d_{\bar{s}_1}^{\dagger}(t_{4-})  d_{s_2}^{\dagger}(t_{5-}) d_{s_2}(t_{6-})  d_{\bar{s}_2}(t_{7-})  d_{\bar{s}_2}^{\dagger}(t_{8-}) } \nonumber \\
    &&+ (-i) \gamma \int dt \sum_{s_1,s_2} \Braket{\mathcal{T}_K d_s(\tau_1) d_s^{\dagger}(\tau_2)  d_{s_1}^{\dagger}(t_{1-}) d_{s_1}(t_{2-})  d_{\bar{s}_1}(t_{3-})  d_{\bar{s}_1}^{\dagger}(t_{4-})  d_{s_2}^{\dagger}(t_{5+}) d_{s_2}(t_{6+})  d_{\bar{s}_2}(t_{7+})  d_{\bar{s}_2}^{\dagger}(t_{8+}) } . \nonumber \\ \label{eqn:perturbation_gamma_1st}
\end{eqnarray}
By applying Wick's theorem to Eq. (\ref{eqn:perturbation_gamma_1st}), we can decompose the first-order corrections into the following four terms:
\begin{eqnarray}
    \int dt \ G_{s,0}^{\mu_1 +}(\tau_1,t) \times (\mathrm{GFs}) \times G_{s,0}^{+ \mu_2}(t,\tau_2) , \label{eqn:classification_selfenergy_gamma_pp} \\
    \int dt \ G_{s,0}^{\mu_1 +}(\tau_1,t) \times (\mathrm{GFs}) \times G_{s,0}^{- \mu_2}(t,\tau_2) , \label{eqn:classification_selfenergy_gamma_pm} \\
    \int dt \ G_{s,0}^{\mu_1 -}(\tau_1,t) \times (\mathrm{GFs}) \times G_{s,0}^{+ \mu_2}(t,\tau_2) , \label{eqn:classification_selfenergy_gamma_mp} \\
    \int dt \ G_{s,0}^{\mu_1 -}(\tau_1,t) \times (\mathrm{GFs}) \times G_{s,0}^{- \mu_2}(t,\tau_2) . \label{eqn:classification_selfenergy_gamma_mm}
\end{eqnarray}
Summing up all contributions from Eq. (\ref{eqn:classification_selfenergy_gamma_pp}), we obtain the $(++)$ component of $\Sigma_{s,\gamma}^{(1)}$.
Similarly, we can calculate $\Sigma_{s,\gamma}^{(1),+-}$, $\Sigma_{s,\gamma}^{(1),-+}$, and $\Sigma_{s,\gamma}^{(1),--}$ by summing up all the contributions from Eqs. (\ref{eqn:classification_selfenergy_gamma_pm}), (\ref{eqn:classification_selfenergy_gamma_mp}), and (\ref{eqn:classification_selfenergy_gamma_mm}), respectively.
As a result, we obtain the self-energy as
\begin{eqnarray}
    &&\Sigma_{s,\gamma}^{(1),++}(t_1,t_2) = \Sigma_{s,\gamma}^{(1),--}(t_1,t_2) \nonumber \\
    &&= \gamma_1 \delta(t_1-t_2) \Bigl[ \frac{1}{2} G_{s,0}^{+-}(t_1,t_1) G_{\bar{s},0}^{-+}(t_1,t_1) G_{\bar{s},0}^{-+}(t_1,t_1)  + \frac{1}{2} G_{s,0}^{-+}(t_1,t_1) G_{\bar{s},0}^{+-}(t_1,t_1) G_{\bar{s},0}^{+-}(t_1,t_1) \nonumber \\
    &&\hspace{2.5cm} + \frac{1}{2} G_{s,0}^{+-}(t_1,t_1) G_{\bar{s},0}^{+-}(t_1,t_1) G_{\bar{s},0}^{+-}(t_1,t_1)  + \frac{1}{2} G_{s,0}^{-+}(t_1,t_1) G_{\bar{s},0}^{-+}(t_1,t_1) G_{\bar{s},0}^{-+}(t_1,t_1) \nonumber \\
    &&\hspace{2.5cm} - G_{s,0}^{+-}(t_1,t_1) G_{\bar{s},0}^{+-}(t_1,t_1) G_{\bar{s},0}^{-+}(t_1,t_1)  - G_{s,0}^{-+}(t_1,t_1) G_{\bar{s},0}^{+-}(t_1,t_1) G_{\bar{s},0}^{-+}(t_1,t_1) \Bigr] ,
\end{eqnarray}
\begin{eqnarray}
    &&\Sigma_{s,\gamma}^{(1),+-}(t_1,t_2)  \nonumber \\
    &&= -\gamma_1 \delta(t_1-t_2) [ G_{s,0}^{+-}(t_1,t_1) G_{\bar{s},0}^{-+}(t_1,t_1) G_{\bar{s},0}^{-+}(t_1,t_1)  +  G_{s,0}^{+-}(t_1,t_1) G_{\bar{s},0}^{+-}(t_1,t_1) G_{\bar{s},0}^{+-}(t_1,t_1) \nonumber \\
    &&\hspace{3.0cm} - 2  G_{s,0}^{+-}(t_1,t_1) G_{\bar{s},0}^{+-}(t_1,t_1) G_{\bar{s},0}^{-+}(t_1,t_1)  ] ,
\end{eqnarray}
and
\begin{eqnarray}
    &&\Sigma_{s,\gamma}^{(1),-+}(t_1,t_2) \nonumber \\
    &&= -\gamma_1 \delta(t_1-t_2) [ G_{s,0}^{-+}(t_1,t_1) G_{\bar{s},0}^{-+}(t_1,t_1) G_{\bar{s},0}^{-+}(t_1,t_1)  +  G_{s,0}^{-+}(t_1,t_1) G_{\bar{s},0}^{+-}(t_1,t_1) G_{\bar{s},0}^{+-}(t_1,t_1) \nonumber \\
    &&\hspace{3.0cm} - 2 G_{s,0}^{-+}(t_1,t_1) G_{\bar{s},0}^{+-}(t_1,t_1) G_{\bar{s},0}^{-+}(t_1,t_1)   ] .
\end{eqnarray}
By replacing the non-interacting GFs with the full GFs and substituting $G_s^{+-}(t,t) = i n_s$ and $G_s^{+-}(t,t) = i (n_s-1) $ into them, we obtain the self-energies shown in Eqs. (\ref{eqn:selfenergy_gamma_1st_ppmm} - \ref{eqn:selfenergy_gamma_1st_RA}).
\section{Detail of the non-unitary Schrieffer-Wolff transformation}

In this section, we provide a detailed derivation of the effective non-Hermitian Kondo model by using the non-unitary Schrieffer-Wolff transformation (SWT).
We first map the density matrix $\rho=\sum_{i_2,i_2} \rho_{i_1,i_2} \Ket{i_1} \Bra{i_2}$ to a vector
\begin{eqnarray}
    \rho' =\sum_{i_2,i_2} \rho_{i_1,i_2} \Ket{i_1} \otimes \Ket{i_2} 
\end{eqnarray}
in a doubled Hilbert space. Here, $\{\ket{i}\}_i$ denotes an orthonormal basis set of the Hilbert space of the system. 
Similarly, Liouvillian $\mathcal{L}$ is mapped to an operator $\mathcal{L}'$ acting on the doubled Hilbert space:
\begin{eqnarray}
    \mathcal{L}' = \sum_{i_1,\cdots,i_4} \mathcal{L}_{i_1 i_2,i_3 i_4} ( \Ket{i_1}\otimes \Ket{i_2} ) (\Bra{i_3} \otimes \Bra{i_4} ) ,
\end{eqnarray}
where $\mathcal{L}_{i_1 i_2,i_3 i_4}=\Bra{i_1}\mathcal{L}(\Ket{i_3}\Bra{i_4})\Ket{i_2}$.
Then, a non-unitary similarity transformation is applied to the Liouvillian and the density matrix as $\tilde{\mathcal{L}} = e^{S} \mathcal{L}' e^{-S}$ and $\tilde{\rho} = e^{S} \rho' $.
The non-Hermitian generator $S$ is defined as
\begin{eqnarray}
    S = S_1 \otimes Q + Q \otimes S_1^{\ast} + S_2 \otimes P + P \otimes S_2^{\ast} ,
\end{eqnarray}
\begin{eqnarray}
    S_i = \sum_{k,s} \sum_{\alpha = p,h } \left( \frac{v_k}{\epsilon_k - \epsilon_i^{\alpha}} n_{d\bar{s}}^{\alpha} c_{ks}^{\dagger} d_s  - \frac{v_k^{\ast}}{\epsilon_k - \epsilon_i^{\alpha}}  d_s^{\dagger} c_{ks} n_{d\bar{s}}^{\alpha} \right) , 
\end{eqnarray}
where $Q=1-P$, $n_{ds}^{p} = 1-n_{ds}^{h} = d_{s}^{\dagger} d_s$, $\epsilon_1^p = (\epsilon_2^p)^{\ast}  =\epsilon_d + U + i \gamma /2$, and $\epsilon_1^h = (\epsilon_2^h)^{\ast} = \epsilon_d - i \gamma /2$.
By expanding the Liouvillian after the transformation up to the second order of $S$ as
\begin{eqnarray}
    \tilde{\mathcal{L}} &\simeq& \mathcal{L}'+[S,\mathcal{L}']+\frac{1}{2}[S,[S,\mathcal{L}']],
\end{eqnarray}
the Liouvillian takes a block-diagonal form with respect to a subspace projected by $P \otimes P + Q \otimes Q$ and its complement:
\begin{eqnarray}
    (P \otimes P + Q \otimes Q)\tilde{\mathcal{L}}(P \otimes Q + Q \otimes P)=(P \otimes Q + Q \otimes P)\tilde{\mathcal{L}}(P \otimes P + Q \otimes Q)=0.
\end{eqnarray}
Note that $P \otimes P + Q \otimes Q + P \otimes Q + Q \otimes P = I \otimes I$, where $I$ is the identity operator. 
The explicit form of the Liouvillian after the non-unitary SWT is given by
\begin{eqnarray}
    i\tilde{\mathcal{L}} &\simeq& P H_{\mathrm{eff}} P \otimes P - P \otimes P H_{\mathrm{eff}}^{\ast} P + Q \tilde{H}_{\mathrm{eff}} Q \otimes Q - Q \otimes Q \tilde{H}_{\mathrm{eff}}^{\ast} Q + \mathcal{L}_{PQ},
    \label{eqn:effective_Liouvillian_full}
\end{eqnarray}
up to the second order of $S$. 
Here,
\begin{eqnarray}
    H_{\mathrm{eff}} &=& \sum_{k,s} \epsilon_{k} c_{ks}^{\dagger} c_{ks} + \sum_s \left( \epsilon_d - \frac{1}{2} \sum_k W_{k,k} \right) d_{s}^{\dagger} d_s  + \sum_{k_1,k_2} J_{k_1,k_2} \bm{S}_{d} \cdot \bm{S}_{k_1k_2} \nonumber \\
    && + \sum_{k_1,k_2} \sum_s \left(\frac{1}{2} W_{k_1,k_2} - \frac{1}{4} J_{k_1,k_2} \right) c_{k_1s}^{\dagger} c_{k_2s} , \label{eqn:effective_Hamiltonian}
\end{eqnarray}
is the effective Hamiltonian in the single-particle subspace, which is the non-Hermitian Kondo model \cite{Nakagawa2018Nov} with a complex-valued spin-exchange coupling
\begin{eqnarray}
    J_{k_1,k_2} &=& v_{k_1} v_{k_2}^{\ast} \Biggl( \frac{1}{\epsilon_{k_1}-\epsilon_2^{h}} + \frac{1}{\epsilon_{k_2}-\epsilon_2^{h}}  - \frac{1}{\epsilon_{k_1}-\epsilon_2^{p}} - \frac{1}{\epsilon_{k_2}-\epsilon_2^{p}} \Biggr) , \label{eqn:complex_Jk1k2}
\end{eqnarray}
and a complex-valued spin-independent potential
\begin{eqnarray}
    W_{k_1,k_2} &=& v_{k_1} v_{k_2}^{\ast} \Biggl( \frac{1}{\epsilon_{k_1}-\epsilon_2^{h}} + \frac{1}{\epsilon_{k_2}-\epsilon_2^{h}} \Biggr) .
\end{eqnarray}
The other effective Hamiltonian
\begin{eqnarray}
    \tilde{H}_{\mathrm{eff}} &=& \sum_{k,s}\epsilon_k c_{ks}^\dag c_{ks}-\sum_{k_1,k_2}J_{k_2,k_1}^*\bm{T}_{d}\cdot\bm{T}_{k_1,k_2}+\sum_{k_1,k_2}\sum_s\Bigl(\frac{1}{2}W_{k_2,k_1}^*-\frac{1}{4}J_{k_2,k_1}^*\Bigr)c_{k_1s}^\dag c_{k_2s} \nonumber \\
    && +\Bigl(2\epsilon_d+U-\sum_kW_{k,k}^*\Bigr)T_d^z+\epsilon_d+\frac{U}{2}-\frac{1}{2}\sum_k W_{k,k}^*, \label{eqn:effective_Hamiltonian_QQ}
\end{eqnarray}
which involves the complex-valued exchange interaction between charge pseudospins
\begin{eqnarray}
    T_{k_1,k_2}^x &=& \frac{1}{2}(c_{k_1\uparrow}^\dag c_{k_2\downarrow}^\dag +c_{k_1\downarrow} c_{k_2\uparrow}),\ 
    T_{k_1,k_2}^y = \frac{1}{2i}(c_{k_1\uparrow}^\dag c_{k_2\downarrow}^\dag -c_{k_1\downarrow} c_{k_2\uparrow}), \ 
    T_{k_1,k_2}^z = \frac{1}{2}(c_{k_1\uparrow}^\dag c_{k_2\uparrow}+c_{k_1\downarrow}^\dag c_{k_2\downarrow}-1),\nonumber\\
    T_d^x &=& -\frac{1}{2}(d_{\uparrow}^\dag d_{\downarrow}^\dag+d_{\downarrow} d_{\uparrow}),\ 
    T_d^y = -\frac{1}{2i}(d_{\uparrow}^\dag d_{\downarrow}^\dag-d_{\downarrow} d_{\uparrow}),\ 
    T_d^z = \frac{1}{2}(d_{\uparrow}^\dag d_{\uparrow}+d_{\downarrow}^\dag d_{\downarrow}-1),\nonumber
\end{eqnarray}
describes the charge Kondo effect in the subspace in which the quantum dot is empty or doubly occupied \cite{Taraphder1991May}. The residual term in Eq.~\eqref{eqn:effective_Liouvillian_full} is expressed as follows:
\begin{eqnarray}
    \mathcal{L}_{PQ} &=& \frac{1}{2}P\Delta SQ\otimes PH_v^*Q-\frac{1}{2}Q\Delta SP\otimes QH_v^*P \nonumber \\ && -\frac{1}{2}PH_vQ\otimes P\Delta S^*Q+\frac{1}{2}QH_vP\otimes Q\Delta S^*P,
\end{eqnarray}
where
\begin{eqnarray}
    \Delta S &=& S_1-S_2 \nonumber \\
    &=& \sum_{k,s}\sum_{\alpha=p,h}\left(\Bigl[\frac{1}{\epsilon_k-\epsilon_1^{\alpha}}-\frac{1}{\epsilon_k-\epsilon_2^{\alpha}}\Bigr]v_kn_{d\bar{s}}^{\alpha}c_{ks}^\dag d_s-\Bigl[\frac{1}{\epsilon_k-\epsilon_1^{\alpha}}-\frac{1}{\epsilon_k-\epsilon_2^{\alpha}}\Bigr]v_k^*d_s^\dag c_{ks}n_{d\bar{s}}^{\alpha}\right),
\end{eqnarray}
and $H_v=\sum_{k,s} v_k ( c_{ks}^{\dagger} d_s + d_s^{\dagger} c_{ks} )$ is the tunneling Hamiltonian.
Thus, the Lindblad equation after the transformation is given by
\begin{eqnarray}
    \frac{d}{dt} \tilde{\rho}_{PP} = ( P \otimes P ) \tilde{\mathcal{L}} ( P \otimes P ) \tilde{\rho}_{PP} +  ( P \otimes P ) \tilde{\mathcal{L}} ( Q \otimes Q ) \tilde{\rho}_{QQ} ,
\end{eqnarray}
where $\tilde{\rho}_{PP}=P \otimes P \tilde{\rho}$ and $\tilde{\rho}_{QQ}=Q \otimes Q \tilde{\rho}$.
In the deep Kondo regime with weak measurement strength, the density-matrix elements are dominated by those in the single-particle subspace $P$.
Hence, we have
\begin{eqnarray}
    \frac{d}{dt} \tilde{\rho}_{PP} \simeq \tilde{\mathcal{L}}_{\mathrm{eff}} \tilde{\rho}_{PP} \label{eqn:porjected_Lindblad}
\end{eqnarray}
where
\begin{eqnarray}
    i \tilde{\mathcal{L}}_{\mathrm{eff}} = P H_{\mathrm{eff}} P \otimes P - P \otimes P H_{\mathrm{eff}}^{\ast} P ,
\end{eqnarray}
is the effective Liouvillian presented in the main text.
The complex-valued spin-exchange coupling constant $J$ at the Fermi level is obtained by setting $\epsilon_{k_1} = \epsilon_{k_2} = 0$ in Eq. (\ref{eqn:complex_Jk1k2}).
Since we are interested in the measurement effect on the Kondo effect, we ignore the spin-independent potential term $W_{k_1,k_2}$ which does not qualitatively change the Kondo physics.
The complex-valued spin-exchange coupling constant at the Fermi level is expanded by the measurement strength $\gamma$ up to the second order as
\begin{eqnarray}
    \tilde{J} &=& \left\{J +  \frac{1}{2} \gamma^2  |v|^2 [  \frac{1}{(\epsilon_d)^3} - \frac{1}{(\epsilon_d+U)^3} ] \right\} + i\gamma |v|^2 \left[ \frac{1}{\epsilon_d^2} + \frac{1}{(\epsilon_d+U)^2} \right]  + O(\gamma^3) .
\end{eqnarray}

We note that the above approximation of neglecting $\tilde{\rho}_{QQ}$ is justified if we postselect the measurement outcomes in which the detector finds a single electron at the quantum dot.
To see this, we rewrite the Lindblad equation as
\begin{eqnarray}
    \frac{d}{dt}\rho &=&-\frac{i}{\hbar}[H,\rho]+\gamma\left(P\rho P-\frac{1}{2}\{ P,\rho\} \right)\nonumber \\
    &=& -\frac{i}{\hbar}[H,\rho]+\gamma\left(Q\rho Q-\frac{1}{2}\{ Q,\rho\} \right).
\end{eqnarray}
Then, when we discard the measurement outcomes that induce quantum jumps by the operator $Q$, the conditional dynamics of the system is described by \cite{Ashida2016Nov, Ashida2018}
\begin{eqnarray}
    \frac{d}{dt}\rho &=&-\frac{i}{\hbar}(H_c\rho-\rho H_c^\dag),
    \label{eqn:postselect_Lindblad}
\end{eqnarray}
where
\begin{eqnarray}
    H_c = H - \frac{i\gamma}{2}Q
\end{eqnarray}
is a non-Hermitian Anderson Hamiltonian that includes the effect of the backaction of postselecting the measurement outcomes. By performing the non-unitary SWT $e^{S_2} H_c e^{-S_2}$, we find that the time evolution equation \eqref{eqn:postselect_Lindblad} after the transformation is equivalent to Eq.~\eqref{eqn:porjected_Lindblad} with the non-Hermitian Kondo model $H_{\mathrm{eff}}$.

At the end of this section, we give a remark on the non-unitary transformation.
The ensemble average of an observable is given by $\Braket{O(t)} = \mathrm{Tr}[O \rho(t)] = \Braket{\rho_T^{\prime},O^{\prime} \rho(t)^{\prime}}$, where $\Braket{\cdot,\cdot}$ denotes the inner product of the doubled Hilbert space, $\rho_T^{\prime} = \sum_i \Ket{i} \otimes \Ket{i}$ is a trace vector, and $O^{\prime}= O \otimes I$ is an observable expressed in the doubled Hilbert space \footnote{One can use another extension $O^{\prime} = I \otimes O$, though there is no difference in the result.}.
Since the non-unitary transformation does not preserve the trace of an operator, the trace vector after the non-unitary SWT is expressed as $\tilde{\rho}_T = M^{-1} \rho_T^{\prime}$ with some non-identity matrix $M$.
Thus, we have $\Braket{O(t)}=\langle\tilde{\rho}_T,\tilde{O}\tilde{\rho}(t)\rangle$, where $\tilde{O} = M^{\dagger} O^{\prime} e^{-S}$ is a transformed observable.
However, when the measurement strength is sufficiently weak, i.e., $\gamma \ll |\epsilon_d|, U$, the non-unitary SWT is almost unitary and therefore we assume $\tilde{O} \simeq O^{\prime}$.
\section{Detailed calculation of spin susceptibility}

In this section, we give a detailed calculation of the spin susceptibility by the perturbation theory with respect to the complex-valued spin-exchange coupling $J$ in the non-Hermitian Kondo model.

\subsection{Perturbation setup}

For later convenience, we consider the non-Hermitian Kondo model
\begin{eqnarray}
    H_{\mathrm{eff}} &=& \tilde{H}_0 + \tilde{H}_1 , 
\end{eqnarray}
where
\begin{eqnarray}
    \tilde{H}_0 &=&  \sum_{k,s} \epsilon_k c_{ks}^{\dagger} c_{ks} + \sum_{s} ( \epsilon_d d_s^{\dagger} d_s + \frac{U}{2} d_s^{\dagger} d_s d_{\bar{s}}^{\dagger} d_{\bar{s}} ) , \label{eqn:Jpert_H0}\\
    \tilde{H}_1 &=& J \sum_{k_1,k_2} \bm{S}_d \cdot \bm{S}_{k_1,k_2} . \label{eqn:Jpert_H1}
\end{eqnarray}
Following the formalism in Ref.~\cite{Seiberer2016Aug}, we define the Keldysh action as
\begin{eqnarray}
    \tilde{S} = \tilde{S}_0 + \tilde{S}_1 ,
\end{eqnarray}
where
\begin{eqnarray}
    \tilde{S}_0 &=& \int_{K} d t \ \Biggl\{ \sum_s d_s^{\dagger}(t) i \partial_t d_s(t) + \sum_{k,s} c_{ks}^{\dagger}(t) i \partial_t c_{ks}(t)  -  \tilde{H}_0(t) \Biggr\} , 
\end{eqnarray}
and
\begin{eqnarray}
    \tilde{S}_1 &=& - \int_{t_i}^{t_f} dt [ \tilde{H}_1(t_{+}) - \tilde{H}_1^{\ast}(t_{-}) ] .
\end{eqnarray}
The spin susceptibility is defined as
\begin{eqnarray}
    \chi_m = \frac{(g \mu_B)^2}{4} \bar{\chi}_{m}(t),
\end{eqnarray}
where
\begin{eqnarray}
    \bar{\chi}_m(t) = i \int_K dt^{\prime} \braket{\sigma_z(t_+) \sigma_z(t^{\prime})} ,
\end{eqnarray}
$g$ is the Land\'{e} g-factor, $\mu_b$ is the Bohr magneton, 
and the operator $\sigma_z$ is defined as
\begin{eqnarray}
    \sigma_z = d_{\uparrow}^{\dagger} d_{\uparrow} - d_{\downarrow}^{\dagger} d_{\downarrow} .
\end{eqnarray}

We note that Wick's theorem cannot be used in the perturbation theory with respect to the spin-exchange coupling since the unperturbed Hamiltonian $\tilde{H}_0$ is not quadratic.
In the following, we calculate the spin susceptibility for arbitrary $\epsilon_d$ and $U$. 
The spin susceptibility of the original non-Hermitian Kondo model [Eq.~(11) in the main text] is obtained by taking the limit of $U=-\epsilon_d\to\infty$ that restricts the occupation number of the dot to unity. 
Note that the parameters $\epsilon_d$ and $U$ in Eq.~\eqref{eqn:Jpert_H0} are introduced for taking this limit and independent of the spin-exchange coupling in Eq.~\eqref{eqn:Jpert_H1}.

\subsection{Unperturbed case}
We first calculate the unperturbed part of the spin susceptibility.
For $J=0$, the dimensionless spin susceptibility is calculated as
\begin{eqnarray}
    \bar{\chi}_m(t) &=& i \int_{t_i}^{t_f} dt^{\prime}_+ \braket{\sigma_z(t_+) \sigma_z(t^{\prime}_+)} 
    - i \int_{t_i}^{t_f} dt^{\prime}_- \braket{\sigma_z(t_+) \sigma_z(t^{\prime}_-)} 
    + i \int_{t_i}^{t_i - i \beta} dt^{\prime}_M \braket{\sigma_z(t_+) \sigma_z(t^{\prime}_M)} \nonumber \\
    &=&  2 \beta \rho_1 ,
\end{eqnarray}
where $\rho_1$ is the probability distribution of the one-particle state, defined as
\begin{eqnarray}
    \rho_1 = \frac{e^{-\beta \epsilon_d}}{1 + 2 e^{-\beta \epsilon_d} + e^{-\beta (2\epsilon_d +U)} } .
\end{eqnarray}
The spin susceptibility for the isolated quantum dot is calculated as
\begin{eqnarray}
    \chi_{m,0} = \frac{(g \mu_B)^2}{2} \beta \rho_1 .
\end{eqnarray}

\subsection{Diverging series}
\begin{figure}[t]
    \centering
    \includegraphics[width=1.00\linewidth]{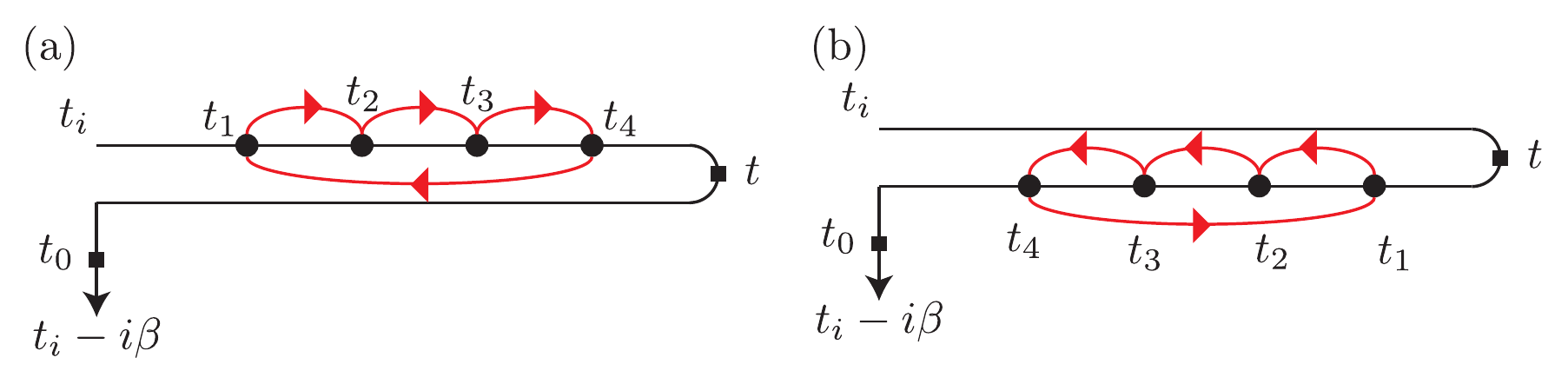}
    \caption{(a) Time-ordered diagram for forward time contour and (b) anti-time-ordered diagram for backward time contour.
    The black arrow denotes the Keldysh contour.
    Solid squares denote the times in which the operator $\sigma_z$ is inserted.
    The complex-valued spin-exchange interaction occurs at the times denoted by solid circles, which are connected by red arrows that express the GFs of the electron reservoir.
    }
    \label{fig:TimeOrderedDiagram}
\end{figure}

Next, we extract the diverging series in the spin susceptibility.
The $N$-th order term of the spin susceptibility is calculated as
\begin{eqnarray}
    i (-i)^N \int_K dt_0 \int_R dt_1 \cdots dt_N \Braket{\mathcal{T}_K \sigma_z(t_0) \sigma_z(t_+) \mathcal{J}(t_1) \cdots \mathcal{J}(t_N) }, \label{eqn:Nth_order_original}
\end{eqnarray}
where $\mathcal{J}(t)$ is an interaction vertex on the Keldysh contour defined as
\begin{eqnarray}
    \mathcal{J}(t_+) = \frac{J}{4} \sum_{s_1,\cdots,s_4} \sum_{k_1,k_2} (\bm{\sigma}_{s_1s_2} \cdot \bm{\sigma}_{s_3s_4} ) d^{\dagger}_{s_1}(t_+) d_{s_2}(t_+) c_{k_1 s_3}^{\dagger}(t_+) c_{k_2 s_4}(t_+) ,
\end{eqnarray}
\begin{eqnarray}
    \mathcal{J}(t_-) = \frac{J^{\ast}}{4} \sum_{s_1,\cdots,s_4} \sum_{k_1,k_2} (\bm{\sigma}_{s_1s_2} \cdot \bm{\sigma}_{s_3s_4} ) d^{\dagger}_{s_1}(t_-) d_{s_2}(t_-) c_{k_1 s_3}^{\dagger}(t_-) c_{k_2 s_4}(t_-) .
\end{eqnarray}
The logarithmic singularity in the spin susceptibility originates from the contributions from the diagrams shown in Fig.~\ref{fig:TimeOrderedDiagram}. 
For example, the diagram shown in Fig.~\ref{fig:TimeOrderedDiagram} (a) represents the contribution
\begin{eqnarray}
    &&(-i) \left( \frac{J}{4}\right)^N \rho_1 \sum_{s_1 \cdots s_N} \sum_{k_1 \cdots k_N} \int_{t_i}^{t_i-i\beta} dt_0 \int_{t_i}^{t} dt_1 \cdots dt_N \Theta(t_1 < \cdots < t_N) \nonumber \\
    &&\times \mathrm{Tr}\left[ \sigma_z (t_0) \sigma_z (t) \bm{\sigma} \cdot \bm{\sigma}_{s_{N-1}s_{N}} (t_N) \cdots \bm{\sigma} \cdot \bm{\sigma}_{s_Ns_1} (t_1)  \right] \nonumber \\
    && \times \bar{g}_{k_1s_1}^{++}(t_1,t_2) \cdots \bar{g}_{k_{N-1}s_{N-1}}^{++}(t_{N-1},t_N) g_{k_Ns_N}^{++}(t_N,t_1) . \label{eqn:all+diagram_after_wick}
\end{eqnarray}
Here, $\Theta(t_1 < \cdots < t_N)$ is a product of the Heaviside unit-step function,
\begin{eqnarray}
    \Theta(t_1 < \cdots < t_N) = \prod_{k=1}^{N-1} \Theta(t_{k+1}-t_k) .
\end{eqnarray}
The time-ordered GF $g_{k_1s_1}^{++}(t_1,t_2)$ of electrons in the electron reservoir is given by
\begin{eqnarray}
    g_{k_1s_1}^{++}(t_1,t_2) &=& (-i) \Braket{\mathcal{T}_K c_{k_1s_1}(t_1) c_{k_1s_1}^{\dagger}(t_2) }_{J=0}  \nonumber \\
    &=& (-i) e^{-i\epsilon_{k_1}(t_1-t_2)} [ \Theta(t_1-t_2) [1-f(\epsilon_{k_1})] - \Theta(t_2-t_1) f(\epsilon_{k_1}) ] .
\end{eqnarray}

The second line in Eq. (\ref{eqn:all+diagram_after_wick}) is calculated by recursively using
\begin{eqnarray}
    \sum_{s_{2c},s_{2d}} (\bm{\sigma}_{s_{3d}s_{2d}} \cdot \bm{\sigma}_{s_{2c}s_{3c}}) (\bm{\sigma}_{s_{2d}s_{1d}} \cdot \bm{\sigma}_{s_{1c}s_{2c}}) = 2 (\bm{\sigma}_{s_{3d}s_{1d}} \cdot \bm{\sigma}_{s_{1c}s_{3c}}) + 2 I_{s_{3d}s_{1d}} I_{s_{1c}s_{3c}}
\end{eqnarray}
as 
\footnote{The time dependence of the left-hand side of Eq.~\eqref{eqn:spin_coef} cancels out because the Pauli operators act only on the single-particle subspace in which the unperturbed states have the same energy. Therefore, the role of time variables in Eq.~\eqref{eqn:spin_coef} is just to specify the time ordering of the operators.}
\begin{eqnarray}
    \mathrm{Tr}\left[ \sigma_z (t_0) \sigma_z (t) \bm{\sigma} \cdot \bm{\sigma}_{s_{N-1}s_{N}} (t_N) \cdots \bm{\sigma} \cdot \bm{\sigma}_{s_Ns_1} (t_1)  \right] = \frac{4}{\sqrt{3}} [(1+\sqrt{3})^{N-1} - (1-\sqrt{3})^{N-1}] . \label{eqn:spin_coef}
\end{eqnarray}

The integrals involving the GFs in Eq.~(\ref{eqn:all+diagram_after_wick}) can be calculated by using the Fourier transformation as
\begin{eqnarray}
    &&\int_{t_i}^{t} dt_1 \cdots dt_N \Theta(t_1 < \cdots < t_N) g_{k_1s_1}^{++}(t_1,t_2) \cdots g_{k_{N-1}s_{N-1}}^{++}(t_{N-1},t_N) g_{k_Ns_N}^{++}(t_N,t_1) \nonumber \\
    &&= \int_{t_i}^{t} dt_1 \int_{t_1}^{t} dt_N \int_{-\infty}^{\infty} dt_2 \cdots dt_{N-1} \bar{g}_{k_1s_1}^{++}(t_1,t_2) \cdots \bar{g}_{k_{N-1}s_{N-1}}^{++}(t_{N-1},t_N) g_{k_Ns_N}^{++}(t_N,t_1) \nonumber \\
    &&= \int \frac{d\omega}{2\pi} \int_{t_i}^{t} dt_1 \int_{t_1}^{t} dt_N g_{k_Ns_N}^{++}(t_N,t_1) e^{-i\omega(t_1-t_N)} \bar{g}_{k_1s_1}^{++}(\omega) \cdots \bar{g}_{k_{N-1}s_{N-1}}^{++}(\omega) \nonumber \\
    &&\simeq - [1-f(\epsilon_{k_N})] [ \partial_{\omega} [\bar{g}_{k_1s_1}^{++}(\omega) \cdots \bar{g}_{k_{N-1}s_{N-1}}^{++}(\omega) ]_{\omega = \epsilon_{k_N}} ,
\end{eqnarray}
where $\bar{g}_{k_1s_1}^{++}(t_1,t_2)$ is the lesser GF 
\begin{eqnarray}
    \bar{g}_{k_1s_1}^{++}(t_1,t_2) &=& i  \Theta(t_2-t_1) e^{-i\epsilon_{k_1}(t_1-t_2)}  f(\epsilon_{k_1}) .
\end{eqnarray}
By replacing the summation over wave numbers with an integral over energy, the logarithmic term is extracted as
\begin{eqnarray}
    &&- \sum_{k_1, \cdots, k_N} [1-f(\epsilon_{k_N})] [ \partial_{\omega} [\bar{g}_{k_1s_1}^{++}(\omega) \cdots \bar{g}_{k_{N-1}s_{N-1}}^{++}(\omega) ]_{\omega = \epsilon_{k_N}} \nonumber \\
    &&= (N-1) \nu^N \int_{-D}^{D} d\epsilon_1 \cdots d\epsilon_N f(\epsilon_{1}) \cdots f(\epsilon_{N-1}) [1-f(\epsilon_N)] \frac{1}{(\epsilon_N-\epsilon_1)^2} \frac{1}{(\epsilon_N-\epsilon_{2})} \cdots \frac{1}{(\epsilon_N-\epsilon_{N-1})} \nonumber \\
    &&\simeq (N-1) \nu^N \int_0^{\beta D} dx_1 \cdots dx_N  \frac{1}{(x_N+x_1)^2} \frac{1}{(x_N+x_{2})} \cdots \frac{1}{(x_N-x_{N-1})} \nonumber \\
    &&= -(N-1) \int_0^{\beta D} dx_N \left[ \frac{1}{x_N+\beta D} - \frac{1}{x_N} \right] \left[ \ln (x_N+\beta D) - \ln (x_N) \right]^{N-2} \nonumber \\
    &&\simeq [\ln \beta D]^{N-1}. \label{eqn:GF_integral}
\end{eqnarray}

By substituting Eqs. (\ref{eqn:spin_coef}) and (\ref{eqn:GF_integral}) into Eq. (\ref{eqn:all+diagram_after_wick}), we have
\begin{align}
    -\frac{1}{\sqrt{3}a} \beta \rho_1 \left( a J \nu \right)^N [ \ln \beta D ]^{N-1} ,
    \label{eqn:Nth_contribution_forward}
\end{align}
where $a = (1+\sqrt{3})/4$.
Summing up Eq.~\eqref{eqn:Nth_contribution_forward} over $N$, we obtain the contribution from the diagrams in Fig.~\ref{fig:TimeOrderedDiagram} (a) as
\begin{eqnarray}
    -\frac{1}{2 \sqrt{3} a} \chi_{m,0} \frac{a J \nu}{1+a J \nu \ln [T/D] }.
\end{eqnarray}

In the same way, the $N$-th order contributions represented by the diagram in Fig.~\ref{fig:TimeOrderedDiagram} (b) can be calculated as following:
\begin{eqnarray}
    &&(-i) \left( -\frac{J^{\ast}}{4}\right)^N \rho_1 \sum_{s_1 \cdots s_N} \sum_{k_1 \cdots k_N} \int_{t_i}^{t_i-i\beta} dt_0 \int_{t_i}^{t} dt_1 \cdots dt_N \Theta(t_1 > \cdots > t_N) \nonumber \\
    &&\times \mathrm{Tr}\left[ \sigma_z (t_0) \bm{\sigma} \cdot \bm{\sigma}_{s_{N-1}s_{N}} (t_N) \cdots \bm{\sigma} \cdot \bm{\sigma}_{s_Ns_1} (t_1)  \sigma_z (t)  \right] \nonumber \\
    && \times g_{k_1s_1}^{--}(t_1,t_2) \cdots g_{k_{N-1}s_{N-1}}^{--}(t_{N-1},t_N) g_{k_Ns_N}^{--}(t_N,t_1) , \label{eqn:all-diagram_after_wick}
\end{eqnarray}
and their summation over $N$ leads to a logarithmic contribution as
\begin{eqnarray}
    -\frac{1}{2 \sqrt{3} a} \chi_{m,0} \frac{a J^{\ast} \nu}{1+a J^{\ast} \nu \ln [T/D] } .
\end{eqnarray}

As a result, we obtain the leading singularity in the spin susceptibility calculated from the diagrams in Fig.~\ref{fig:TimeOrderedDiagram} as
\begin{eqnarray}
    \chi_m \sim \chi_{m,0} \left[1 - \frac{1}{\sqrt{3}a} \mathrm{Re}\left[ \frac{a J \nu}{1+a J \nu \ln [T/D] } \right]  \right] .
\end{eqnarray}
In the main text, we subtracted the unperturbed susceptibility from this result and wrote the singular contribution as $\tilde{\chi}_m=\chi_m-\chi_{m,0}$.

\section{Time evolution under the projective measurement}

In the main text, we interpreted the suppression of the Kondo effect due to the measurement backaction by using the unitary SWT. 
In this section, we show detailed calculations of the unitary SWT and discuss time evolution of the density matrix under projective measurement.
Then, we show that the measurement backaction in the frame after the uniraty SWT projects a single-particle state of the quantum dot onto another state with charge fluctuations, thereby weakening the Kondo effect.

The unitary SWT is widely used to map the Anderson impurity model to the Kondo model \cite{Schrieffer1966Sep}.
Here, for simplicity, we consider a limit in which the excitation energies of the quantum dot are sufficiently larger than those in the electron reservoirs, i.e., $\epsilon_k \ll |\epsilon_d|, |\epsilon_d + U|$.
In this limit, the generator of the unitary SWT is given by
\begin{eqnarray}
    S_{\mathrm{u}} \simeq  v \sum_s \left[ \frac{1}{\epsilon_d+U} d^{\dagger}_{\bar{s}} d_{\bar{s}} (d_{s}^{\dagger} c_{0s} - c_{0s}^{\dagger} d_{s}) + \frac{1}{\epsilon_d} d_{\bar{s}} d_{\bar{s}}^{\dagger} (d_{s}^{\dagger} c_{0s} - c_{0s}^{\dagger} d_{s}) \right] ,
\end{eqnarray}
where $c_{0s} = \sum_k c_{ks}$ is the annihilation operator of an electron at the site next to the quantum dot.
Through this transformation, a single-particle state of the quantum dot is mapped to a state with charge fluctuations. For example, we have
\begin{eqnarray}
    e^{S_{\mathrm{u}}} \Ket{\uparrow \downarrow}_{\mathrm{A}} = M_{\uparrow\downarrow,\uparrow\downarrow}  \Ket{\uparrow \downarrow}_{\mathrm{K}} + M_{\uparrow\downarrow,\downarrow \uparrow}  \Ket{\downarrow \uparrow}_{\mathrm{K}} +  M_{\uparrow\downarrow,20}  \Ket{20}_{\mathrm{K}} +  M_{\uparrow\downarrow,02}  \Ket{02}_{\mathrm{K}} ,
\end{eqnarray}
where 
\begin{eqnarray}
    M_{\uparrow\downarrow,\uparrow\downarrow} &=& 1 - \frac{v^2}{2} [ \frac{1}{\epsilon_d^2} + \frac{1}{(\epsilon_d+U)^{2}} ] + O(v^3), \nonumber \\
    M_{\uparrow\downarrow,\downarrow \uparrow} &=& - \frac{v^2}{2} [ \frac{1}{\epsilon_d^2} + \frac{1}{(\epsilon_d+U)^{2}} ] + O(v^3), \nonumber \\
    M_{\uparrow\downarrow,20} &=&  \frac{v}{\epsilon_d+U} + O(v^3), \nonumber \\
    M_{\uparrow\downarrow,02} &=& -\frac{v}{\epsilon_d}  + O(v^3) .
\end{eqnarray}
Here, $\Ket{\sigma}$, $\Ket{0}$, and $\Ket{2}$ denote a singly occupied state with spin $\sigma$, an empty state, and a doubly occupied state, respectively, and $\Ket{ij}_{\mathrm{A}}$ ($\Ket{ij}_{\mathrm{K}}$) denotes a quantum state in the frame before (after) the transformation, where the dot site is in the $\Ket{i}$ state and the reservoir site next to the dot is in the $\Ket{j}$ state.
Thus, the backaction of the projective measurement of the single-particle states produces charge fluctuations at the quantum dot in the frame after the transformation.

Having obtained the basis after the unitary SWT, we now calculate an infinitesimal time evolution under the projective measurement.
Here we focus on the probability distribution of several charge configurations such as
\begin{eqnarray}
    && P_{\uparrow \downarrow}(t) = {}_{\mathrm{K}}\Bra{\uparrow \downarrow} \rho(t) \Ket{\uparrow \downarrow}_{\mathrm{K}} , P_{\downarrow \uparrow}(t) = {}_{\mathrm{K}}\Bra{\downarrow \uparrow} \rho(t) \Ket{\downarrow \uparrow}_{\mathrm{K}} , \nonumber \\
    && P_{20}(t) = {}_{\mathrm{K}}\Bra{20} \rho(t) \Ket{20}_{\mathrm{K}} , P_{02}(t) = {}_{\mathrm{K}}\Bra{02} \rho(t) \Ket{02}_{\mathrm{K}} . 
\end{eqnarray}
We consider an initial state
\begin{eqnarray}
    \rho(t_i) = \Ket{02}_{\mathrm{K}\ \mathrm{K}}\Bra{02} \rho_0 + \left[ \Ket{\uparrow \downarrow}_{\mathrm{K}\ \mathrm{K}}\Bra{\uparrow \downarrow} + \Ket{\downarrow \uparrow}_{\mathrm{K}\ \mathrm{K}}\Bra{\downarrow \uparrow} \right] \rho_1 +  \Ket{20}_{\mathrm{K}\ \mathrm{K}}\Bra{20} \rho_2,
\end{eqnarray}
where $\rho_0 + 2 \rho_1 + \rho_2 = 1$.
The initial state shown in the main text is a special case of this state and is obtained by setting $\rho_0=\rho_2=0$. 
Then, the initial values of the probability distribution read
\begin{eqnarray}
     \quad  P_{02}(t_i) = \rho_0 , \quad P_{\uparrow \downarrow}(t_i) = P_{\downarrow \uparrow}(t_i) = \rho_1, \quad P_{20}(t_i) = \rho_2 . 
\end{eqnarray}
The infinitesimal time evolution of the probability distribution is calculated from the Lindblad equation:
\begin{eqnarray}
    \frac{d}{dt} P_{ij}(t_i) ={}_{\mathrm{K}}\Bra{ij} \left[ \gamma\tilde{L} \rho(t_i) \tilde{L}^{\dagger} - \frac{\gamma}{2} \{ \tilde{L}^{\dagger}\tilde{L}, \rho(t_i) \}\right] \ket{ij}_{\mathrm{K}},
\end{eqnarray}
where $\tilde{L} = e^{S_{\mathrm{u}}} L e^{-S_{\mathrm{u}}}$.
The time evolution equations are approximated up to the second order of $v$ as
\begin{eqnarray}
    \frac{d}{dt} P_{02}(t_i) &=& 2 \gamma \frac{v^2}{\epsilon_d^2} (\rho_1-\rho_0) + O(v^3), \nonumber \\
    \frac{d}{dt} P_{\uparrow \downarrow}(t_i) &=& \frac{d}{dt} P_{\downarrow \uparrow }(t_i) =  \gamma \left[ \frac{v^2}{\epsilon_d^2} (\rho_0-\rho_1) \frac{v^2}{(\epsilon_d+U)^2} (\rho_2-\rho_1) \right] + O(v^3), \nonumber \\
    \frac{d}{dt} P_{20}(t_i) &=& 2 \gamma \frac{v^2}{(\epsilon_d+U)^2} (\rho_1-\rho_2) + O(v^3) .
\end{eqnarray}
Since $\rho_1$ is typically larger than $\rho_0$ and $\rho_2$ in the Kondo regime, we conclude that the distribution of the charge-transferred states increases, $\frac{d}{dt} P_{02}(t_i) , \frac{d}{dt} P_{20}(t_i) > 0$, and the distribution of the single-particle states decreases, $\frac{d}{dt} P_{\uparrow \downarrow}(t_i) , \frac{d}{dt} P_{\downarrow \uparrow }(t_i) < 0$.
This tendency suggests that the projective measurement of the singly occupied states facilitates charge fluctuations in the frame after the unitary SWT, which prevent the formation of the Kondo singlet state.

\bibliography{references}